\documentclass[longauth]{aa}
\usepackage[varg]{txfonts}

\begin{document}

\title{Initial LOFAR observations of  epoch of reionization windows:\\ II. Diffuse polarized emission in the ELAIS-N1 field} 
\titlerunning{Initial LOFAR-EoR observations - diffuse polarized emission in the ELAIS-N1 field}

\author{V. Jeli\'{c}\inst{1,2}\thanks{E-mail:vjelic@astro.rug.nl} \and A.~G. de Bruyn\inst{1,2}  \and M. Mevius\inst{1} \and 
F.~B. Abdalla\inst{3} \and K.~M.~B. Asad\inst{1} \and G. Bernardi\inst{4} \and M.~A. Brentjens\inst{2} \and S. Bus\inst{1} \and E. Chapman\inst{3} \and B. Ciardi\inst{5} \and S. Daiboo\inst{1}
% ...
}
\authorrunning{V. Jeli\'{c} et al.}

\institute{Kapteyn Astronomical Institute, University of Groningen, PO Box 800, 9700 AV Groningen, the Netherlands
  \and ASTRON - the Netherlands Institute for Radio Astronomy, PO Box 2, 7990 AA Dwingeloo, the Netherlands
  \and Department of Physics \& Astronomy, University College London, Gower Street, London WC1E 6BT, UK
  \and SKA SA, 3rd Floor, The Park, Park Road, Pinelands, 7405, South Africa
  \and Max-Planck Institute for Astrophysics, Karl-Schwarzschild-Strasse 1, D-85748 Garching bei M\"unchen, Germany
% ...
}

\date{Received 15/04/2014 / Accepted 07/07/2014}

\abstract {} 
{This study aims to characterise the polarized foreground emission in the ELAIS-N1 field and to address its possible implications for extracting of the cosmological 21-cm signal from the LOw-Frequency ARray -  Epoch of Reionization (LOFAR-EoR) data.}
{We used the high band antennas of LOFAR to image this region and RM-synthesis to unravel structures of polarized emission at high Galactic latitudes.}
{The brightness temperature of the detected Galactic emission is on average $\sim4~{\rm K}$ in polarized intensity and covers the range from $-10$ to $+13~{\rm rad~m^{-2}}$ in Faraday depth. The total polarized intensity and polarization angle show a wide range of morphological features. We have also used the Westerbork Synthesis Radio Telescope (WSRT)  at 350~MHz to image the same region. The LOFAR and WSRT images show a similar complex morphology at comparable brightness levels, but their spatial correlation is very low. The fractional polarization at 150~MHz, expressed as a percentage of the total intensity, amounts to $\approx1.5\%$.  There is no indication of diffuse emission in total intensity in the interferometric data, in line with results at higher frequencies}
{The wide frequency range, high angular resolution, and high sensitivity make LOFAR an exquisite instrument for studying Galactic polarized emission at a resolution of $\sim1-2~{\rm rad~m^{-2}}$ in Faraday depth. The different polarized patterns observed at 150~MHz and 350~MHz are consistent with different source distributions along the line of sight wring in a variety of Faraday thin regions of emission. The presence of polarized foregrounds is a serious complication for  epoch of reionization experiments.  To avoid the leakage of polarized emission into total intensity, which can  depend on frequency, we need to calibrate the instrumental polarization across the field of view to a small fraction of $1\%$.}


\maketitle 

\section{Introduction}\label{sec:intro}
The LOw-Frequency ARray -  Epoch of Reionization (LOFAR-EoR) key science project will use the LOFAR radio telescope to study the  epoch of reionization \citep[][]{haarlem13}. The EoR is a pivotal period in the history of the Universe during which the all-pervasive cosmic gas was transformed from a neutral to an ionized state. It  holds the key to structure formation and the evolution of the Universe as we know it today, and touches upon fundamental questions in cosmology.

The LOFAR-EoR project plans to probe the EoR in up to five observing fields (de Bruyn et al, in prep.).  Three of these fields have been observed during the commissioning phase of LOFAR. The first field is centred on the North Celestial Pole (NCP). The second field coincides with the ELAIS-N1 field, while the third contains a very bright source, 3C196. The choice of these three fields was motivated by the desire to address most of the problems and challenges that will affect much longer LOFAR-EoR observations. 

The results of our commissioning observations are presented in two papers. In the first paper \citep{yatawatta13a} we tested various calibration approaches and conducted a thorough analysis of the noise by analysing the NCP observations. The NCP field only has a few bright sources, and the diffuse linearly polarized emission from our Galaxy is relatively faint. In the NCP data  we reached a noise level of about $100~{\mu \rm Jy~PSF^{-1}}$ (PSF: point spread function) with then still poorly calibrated LOFAR array. 

In this paper we present LOFAR observations of the ELAIS-N1 field, which was found to have bright polarized emission coming from the Galactic foreground.  The ELAIS-N1 field is one of the northern fields of the European Large Area {\em Infrared Space Observatory} (ISO) Survey. It is a field with no radio sources brighter than $3~{\rm Jy}$ at $325~{\rm MHz}$ \citep[WENSS survey;][]{rengelink97}. Given its location in the Galactic halo, we do not expect high levels of emission from our Galaxy in total intensity. However,  linear polarization at surface brightness levels of a few K has been detected at $350~{\rm MHz}$ (PI: V. Jeli\'{c}) using the Westerbork Synthesis Radio Telescope (WSRT). The  emission is confined to Faraday depths varying from $-10~{\rm rad~m^{-2}}$ to $+10~{\rm rad~m^{-2}}$, and they show large-scale spatial Faraday depth gradients of a few ${\rm rad~m^{-2}~deg^{-1}}$.  This field is therefore suitable for polarimetric studies of Galactic foreground emission and of its possible contaminating effects on the feeble cosmological signals coming from the EoR.

The ELAIS-N1 field will also be targeted with Subaru Hyper Suprime-Cam \citep{miyazaki02} to search for high-redshift Ly$\alpha$ emitters (LAEs). With a significant number of detected LAEs one could  in principle study reionization using both the shape and normalization of the cross-power spectrum between the galaxies and EoR \citep[e.g.][]{wiersma13}. Thus, the multi-wavelength aspect of this field will play an important role in the detection of the cosmological signal by providing further insight into physical processes during the EoR.

This paper is organised as follows. In Section 2 we give an overview of the observational setup and the data reduction. The initial widefield images of the ELAIS-N1 field in total intensity and polarization are presented in Section 3, where we also present the effect of correction for Faraday rotation in the Earth's ionosphere. In Section 4 we discuss the properties of the detected polarized emission and importance to understand it for a proper extraction of the cosmological signal from the LOFAR-EoR data. We summarise and conclude in Section 5.

\section{Observation and data reduction}\label{sec:data}
The ELAIS-N1 field was observed on 2 and 3 July 2011 with 55 LOFAR High Band Antenna (HBA) stations in the Netherlands.
The observation was part of an initial suite of LOFAR commissioning observations. The array configuration consisted of 46 core stations (CS)  and nine remote stations (RS). The phase centre was set at RA 16h 14m and Dec +54d 30m (J2000). Data were recorded 
in the frequency range from $138~{\rm MHz}$ to $185~{\rm MHz}$ distributed over 240 sub-bands. Each sub-band has a width of $195~{\rm kHz}$ covered by 64 channels. The total integration time was $7~{\rm h}$ (mostly night time) with a correlator integration time of $2~{\rm s}$. The \emph{uv} coverage was fully sampled up to baselines of $2~{\rm km}$.  We refer to \citet{haarlem13} for a detailed overview of the LOFAR radio telescope and its frequency characteristics. 

The initial pre-processing was done on the CEP2 cluster by the Radio Observatory of ASTRON (the Netherlands Institute of Radio Astronomy). All other processing was done on a CPU/GPU\footnote{CPU: Central Processing Unit; GPU: Graphics Processing Unit} cluster dedicated to the LOFAR-EoR project, at the University of Groningen, the Netherlands. During the processing we used the LOFAR-EoR Diagnostic Database (LEDDB) that is used for the storage, management, processing, and analysis of the LOFAR-EoR observations  \citep{martinezrubi13}.

\begin{figure}
\centering \includegraphics[width=.45\textwidth]{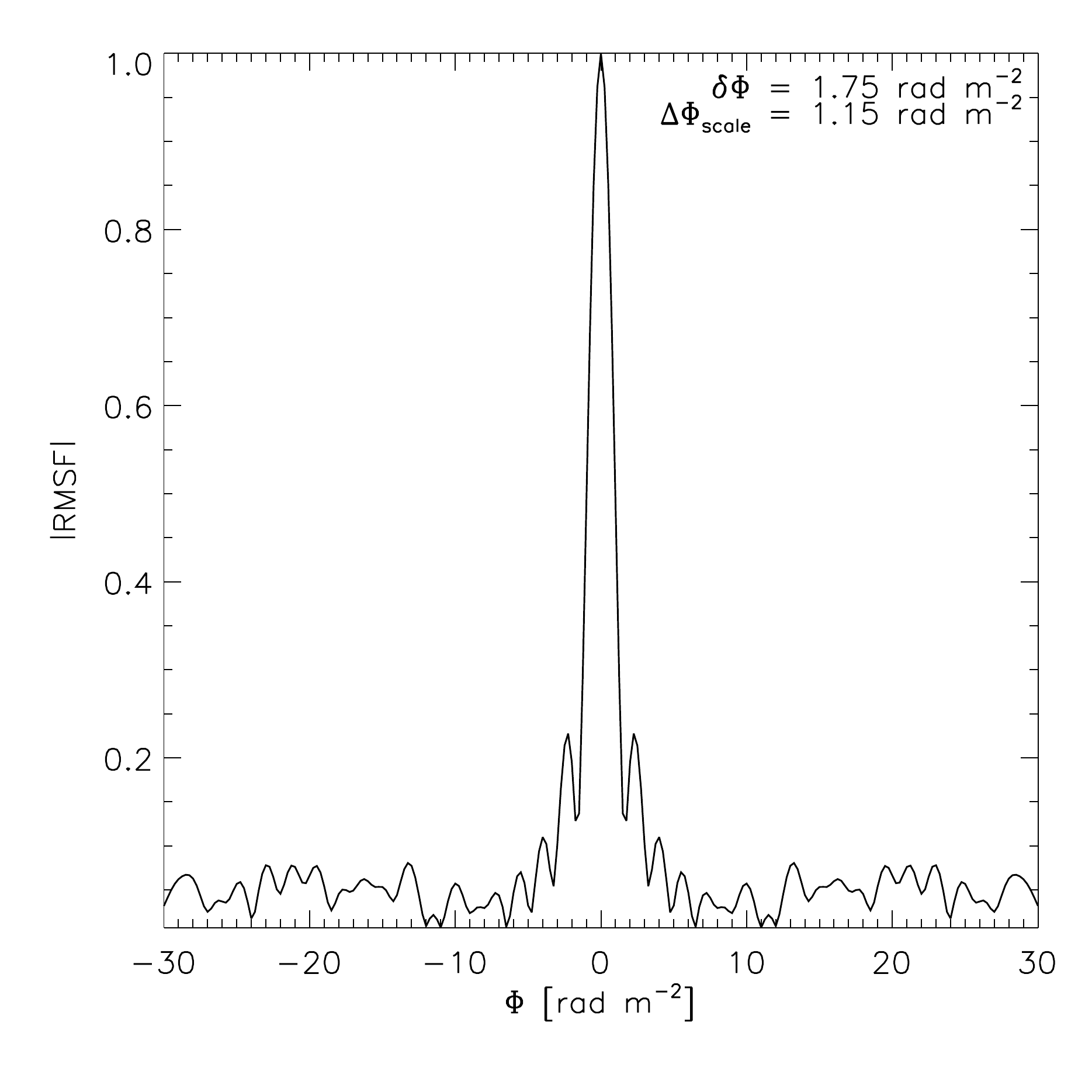}
\caption{Rotation measure spread function (RMSF) for the ELAIS-N1 observation. A resolution in Faraday depth space is $\delta\Phi=1.75~{\rm rad~m^{-2}}$, while the largest Faraday structure that can be reliably detected has a width of  $\Delta\Phi_{\rm scale}=1.15~{\rm rad~m^{-2}}$.}
\label{fig:rmsf}
\end{figure}

\subsection{Initial pre-processing (flagging and averaging)}
The LOFAR observing frequencies are affected by man-made radio frequency interference \citep[RFI;][]{offringa13}.  RFI mitigation works best on data with a very fine resolution in time and frequency. Therefore, the first step in our initial pre-processing is flagging of the data using the \texttt{aoflagger} \citep{offringa10,offringa12}.  On average about $3\%-4\%$ of our data were flagged. However,  around three frequencies ($170~{\rm MHz}$, $178~{\rm MHz}$, and $182~{\rm MHz}$) the percentage of RFIs is much higher ($>30\%$). The second frequency corresponds to the $\sim1.5~{\rm MHz}$ wide Digital Audio Broadcasting (DAB) band C allocated in the Netherlands. A total of  six stations were not delivering good data at the time of our observation and were not  included in the subsequent processing. The 4 edge channels of the 64 channel sub-band  are flagged to remove edge effects from the polyphase filter. After flagging, the data are averaged to 15 channels per sub-band to reduce the data volume for further processing. 
 
\begin{figure*}[ptb]
\centering \includegraphics[width=.75\textwidth]{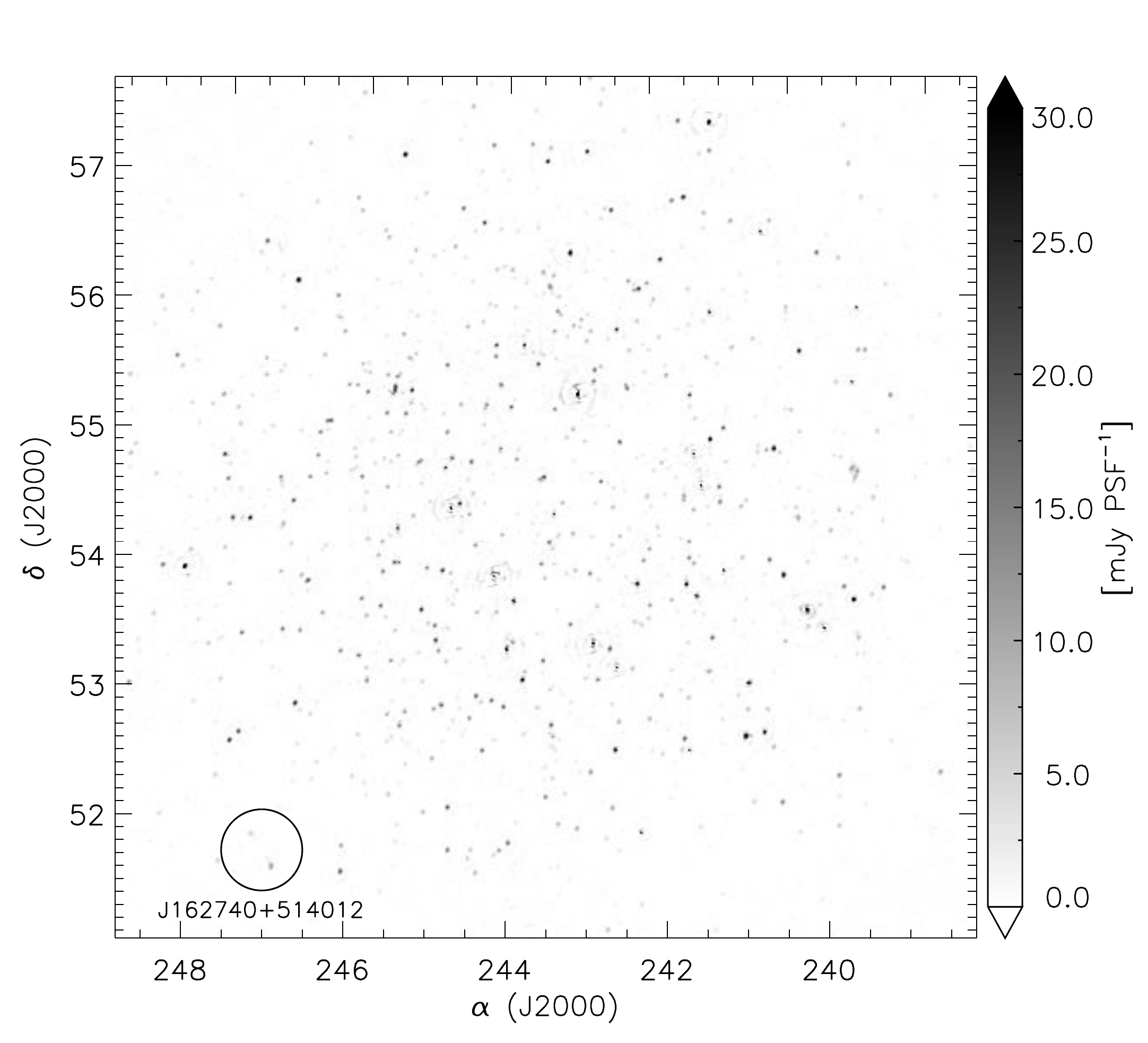}
\caption{Frequency-averaged Stokes I image of the ELAIS-N1 region.  The image is $6.6^\circ\times6.6^\circ$ in size, with a PSF of $16.0''\times8.8''$, and the noise level is $1.0~{\rm mJy~PSF^{-1}}$. We are a factor of $\sim 2 $ above the thermal noise.}
\label{fig:stokesI}
\end{figure*}
 
\subsection{Sky model}
The sky model used for the initial calibration of the ELAIS-N1 field contains approximately 30 of the brightest discrete sources. The flux and spectral index of these sources are determined from the WENSS\footnote{The Westerbork Northern Sky Survey, http://www.astron.nl/wow/} \citep{rengelink97} and VLSS\footnote{The VLA Low-Frequency Sky Survey, http://lwa.nrl.navy.mil/VLSS/} \citep{cohen07}  radio source catalogues at $325~{\rm MHz}$ and $74~{\rm MHz}$.   

After the initial calibration and imaging, we update the sky model by extracting the source information from the LOFAR images themselves. For this we use \texttt{Duchamp} \citep[v. 1.1.11;][]{whiting12}, a source finder that creates masks around potential sources, then we used \texttt{buildsky} \citep[v. 0.0.5;][]{yatawatta13a} to create a model with the minimum number of required source components. Our updated sky model has $\sim200$ sources and assumes that all sources are unpolarized.  This assumption has no effect on the calibration. The diffuse polarized emission is not part of the sky model.  

\subsection{Calibration and source subtraction}\label{sec:cal}
The main steps in the calibration and source subtraction are very similar to the steps presented in our first paper \citep{yatawatta13a}. Therefore we limit ourselves here to a brief overview.

We begin with a direction-independent calibration to correct for clock errors and ionospheric errors effecting the brightest sources in the image. This is done separately on each sub-band using the \texttt{Black Board Selfcal} (\texttt{BBS}) package \citep{pandey09}. Each sub-band has 15 channels of 12~kHz at $2~{\rm s}$ integration but we determine calibration solutions per sub-band for every $10~{\rm s}$. The data are also corrected for the element and station beam gains for the centre of the image.

After performing direction-independent calibration, the corrected data are flagged for bad solutions and averaged to one  channel per 180~kHz sub-band and $10~{\rm s}$ integration time.  Direction-dependent station beam and ionospheric corrections for the brightest sources were determined using  \texttt{SAGECal}, which is based on Expectation Maximization \citep{yatawatta08, kazemi11, yatawatta13b, kazemi13}. We subtract around 200 sources within the image, clustered in $\sim$50 different directions within the main field of view. The very  bright A-team sources (CasA and CygA) and a few bright 3C-sources located near to  the ELAIS-N1 field (e.g. 3C295) were among the 50 clusters.

For the purpose of this work we use data calibrated with \texttt{SAGECal} to build and update our sky model.  The analysis  of diffuse polarized emission uses the data calibrated only with \texttt{BBS}.  \texttt{SAGECal} calibration will suppress large scale diffuse emission, because this emission is not part of the sky model. Since our sky model only contains discrete sources,  we do not use  \texttt{SAGECal} calibration for polarimetric study of diffuse emission. The dynamic range in the polarized emission also does not require sophisticated calibration.  

\subsection{Imaging}
For imaging we use \texttt{AWimager}. \texttt{AWimager} is a fast imager developed and optimized for LOFAR \citep{tasse13}. It is based on full-polarization A-projection that can deal with non-coplanar arrays, arbitrary station beams, and non-diagonal Mueller matrices. The algorithm is designed to correct for all direction-dependent effects varying in time and frequency, including individual station and dipole beams, the projection of the dipoles on the sky and the beam forming, as well as ionospheric refraction effects.

To update the sky model, we make images in total intensity of calibrated and source-subtracted data. In order to create accurate source models, these images have the highest resolution available at the time the data were taken ($\sim 20''$). We use uniform weighting and the subtracted sources are restored onto these images, after convolving with the nominal Gaussian PSF. We also made very large images with low angular resolution to identify bright radio sources surrounding the ELAIS-N1 field,  whose sidelobes contaminate the emission in the inner area.  To analyse the polarization of the diffuse Galactic emission, which mostly appears on spatial scales greater than a few arcmin, we make a lower resolution images in all Stokes parameters (IQUV). These images are produced using only baselines smaller than 1000 wavelengths, providing a frequency independent resolution of about 3 arc min. We used robust (Briggs) weighting with robustness parameter equal to 0. 

The beam pattern of a LOFAR-HBA station can be described as the product of an antenna beam pattern and the array beam pattern of the station \citep{haarlem13}. To first order (i.e. excluding mutual coupling) the station (array) beam is scalar, has no polarizing characteristics itself and depends mainly on the  geometry of the tile distribution.  The element beam pattern is strongly polarized \citep{hamaker06}. Its polarization response is related to the projection of the beam patterns of two orthogonal dipoles on the sky and the changing parallactic angle. During a long synthesis observation, spurious polarization is produced by the field rotation relative to the dipoles  as well as  by the movement of the station beam through the polarized pattern of the averaged beam of the element antennas \citep[for a detailed discussion we refer to][]{bregmanPhD}.

We correct the data for the beam pattern in two steps. The first correction is applied during the calibration, using \texttt{BBS}. The data are corrected for both the array and the element beam gain at the centre of the image. The relative variation of the element beam pattern across the field of view, as well as the temporal changes are taken into account during the imaging step, using the \texttt{AWimager}.

\subsection{Rotation Measure synthesis}\label{sec:RMsyn}
The technique of Rotation Measure (RM) synthesis \citep{brentjens05} is used to unravel the linearly polarized emission as a function of Faraday depth ($\Phi$). The Faraday depth is defined as:
\begin{equation}\label{eq:FD}
\frac{\Phi}{[\rm rad~m^{-2}]}=0.81\int_{source}^{observer}\frac{n_e}{[\rm cm^{-3}]}\frac{B_{\parallel}}{\rm [\mu G]}\frac{{\rm d}l}{\rm [pc]},
\end{equation}
where $n_e$ is electron density; $B_{\parallel}$ is the magnetic field component parallel to the line of sight ${\rm d}l$ and the integral is taken over the entire path from the source to the observer.  A positive Faraday depth implies a magnetic field component pointing towards the observer and a negative Faraday depth implies a magnetic field component pointing away from the observer.

The RM synthesis technique takes advantage of the relationship, which exists between the measured complex polarization $P(\lambda^2)=Q(\lambda^2)+iU(\lambda^2)$ in $\lambda^2$-space and Faraday depth:
\begin{equation}\label{eq:RMsynth}
F(\Phi)=\frac{1}{W(\lambda^2)}\int_{-\infty}^{+\infty} P(\lambda^2)e^{-i2\Phi\lambda^2}{\rm d}\lambda^2,
\end{equation} 
where $W(\lambda^2)$ is the sampling function, also known as the rotation measure spread function (RMSF). Note that we can only sample a finite positive range of wavelengths, resulting in an incomplete $F(\Phi)$. The RM synthesis method is constrained  by the spectral bandwidth ($\Delta\lambda^2$), the spectral resolution ($\delta\lambda^2$), and the minimum ($\lambda_{\rm min}^2$) of the measured $\lambda^2$ distribution. These observational parameters are also directly linked to three physical quantities in Faraday depth space: (i) the maximum detectable Faraday depth, $\Phi_{\rm max}\approx\sqrt{3}/\delta\lambda^2$; (ii) the largest structure that can be resolved in Faraday depth, $\Delta\Phi_{\rm scale}\approx\pi/\lambda_{\rm min}^2$; and (iii) the resolution in Faraday depth space, $\delta\Phi\approx2\sqrt{3}/\Delta\lambda^2$, which defines the minimum separation between two different structures that are detectable. 

For this work we are using the RM synthesis code written by M. Brentjens and we apply it to $\sim 200$ sub-bands, which have comparable noise level. We first synthesized a low resolution Faraday cube over a wide range in Faraday depth to determine where polarized emission could be detected. The final cube covers a Faraday depth range from $-30~{\rm rad~m^{-2}}$ to $+30~{\rm rad~m^{-2}}$ with $0.25~{\rm rad~m^{-2}}$ step. The absolute value of the RMSF corresponding to the frequency coverage of the observation is given in Fig.~\ref{fig:rmsf}. The resolution in Faraday depth space is $\delta\Phi=1.75~{\rm rad~m^{-2}}$, while the largest Faraday structure that can be resolved is $\Delta\Phi_{\rm scale}=1.15~{\rm rad~m^{-2}}$. Since the resolution is higher than the maximum detectable scale, we can only detect Faraday thin structures. A structure is Faraday thin if $\lambda^2\Delta\Phi\ll1$, where $\Delta\Phi$ denotes the extent of the structure in Faraday depth. If $\lambda^2\Delta\Phi\gg1$, then the structure is called Faraday thick. 

\begin{figure}[tb]
\centering \includegraphics[width=.45\textwidth]{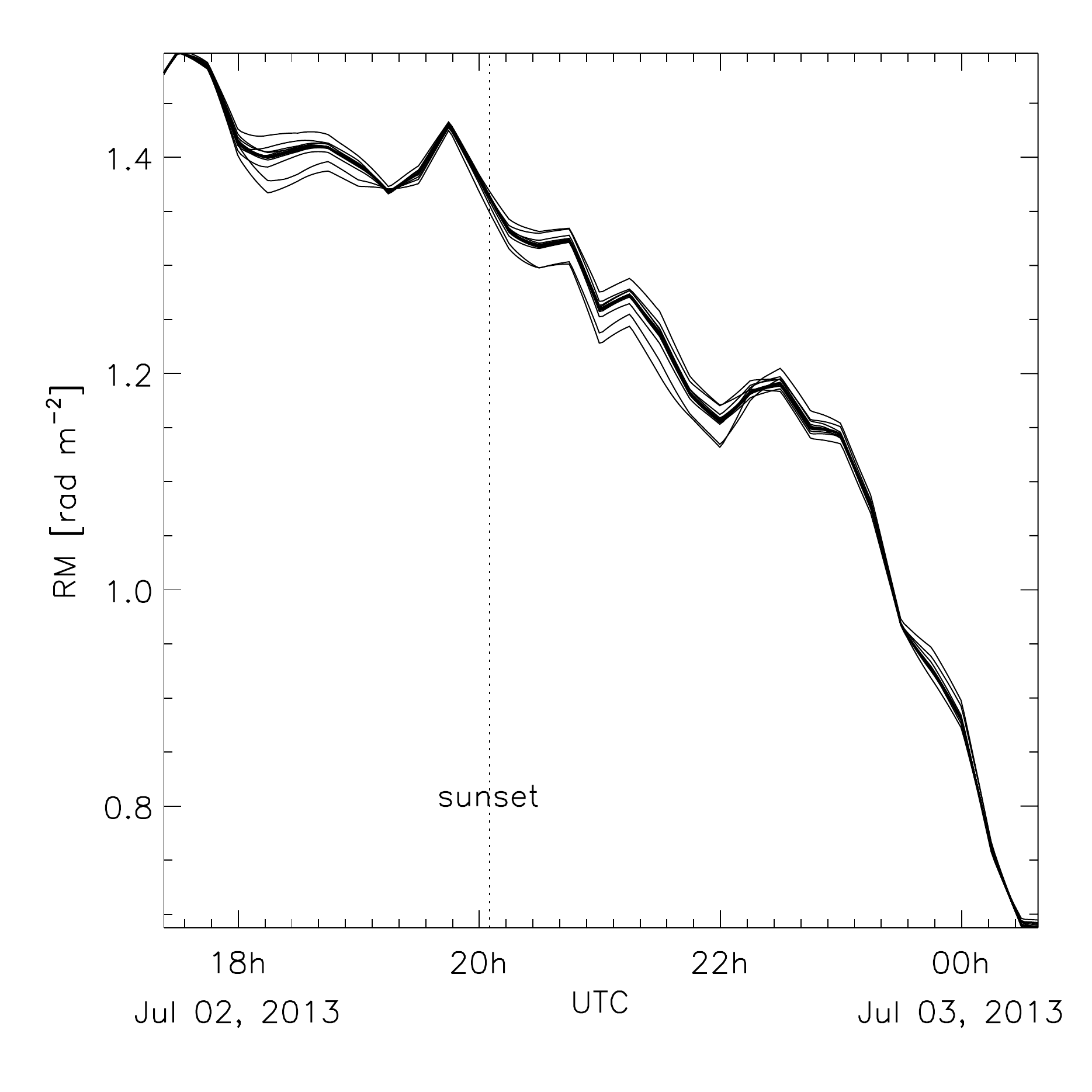}
\caption{Ionospheric RM values for different LOFAR-HBA stations during the ELAIS-N1 observation estimated from Global Ionospheric Maps (GIMs). Note that GIMs have an error of 1~TEC unit, which translates to an RM error of  $0.1~{\rm rad~m^{-2}}$.}
\label{fig:ionRM}
\end{figure}

\begin{figure*}[ptb]
\centering \includegraphics[width=.86\textwidth]{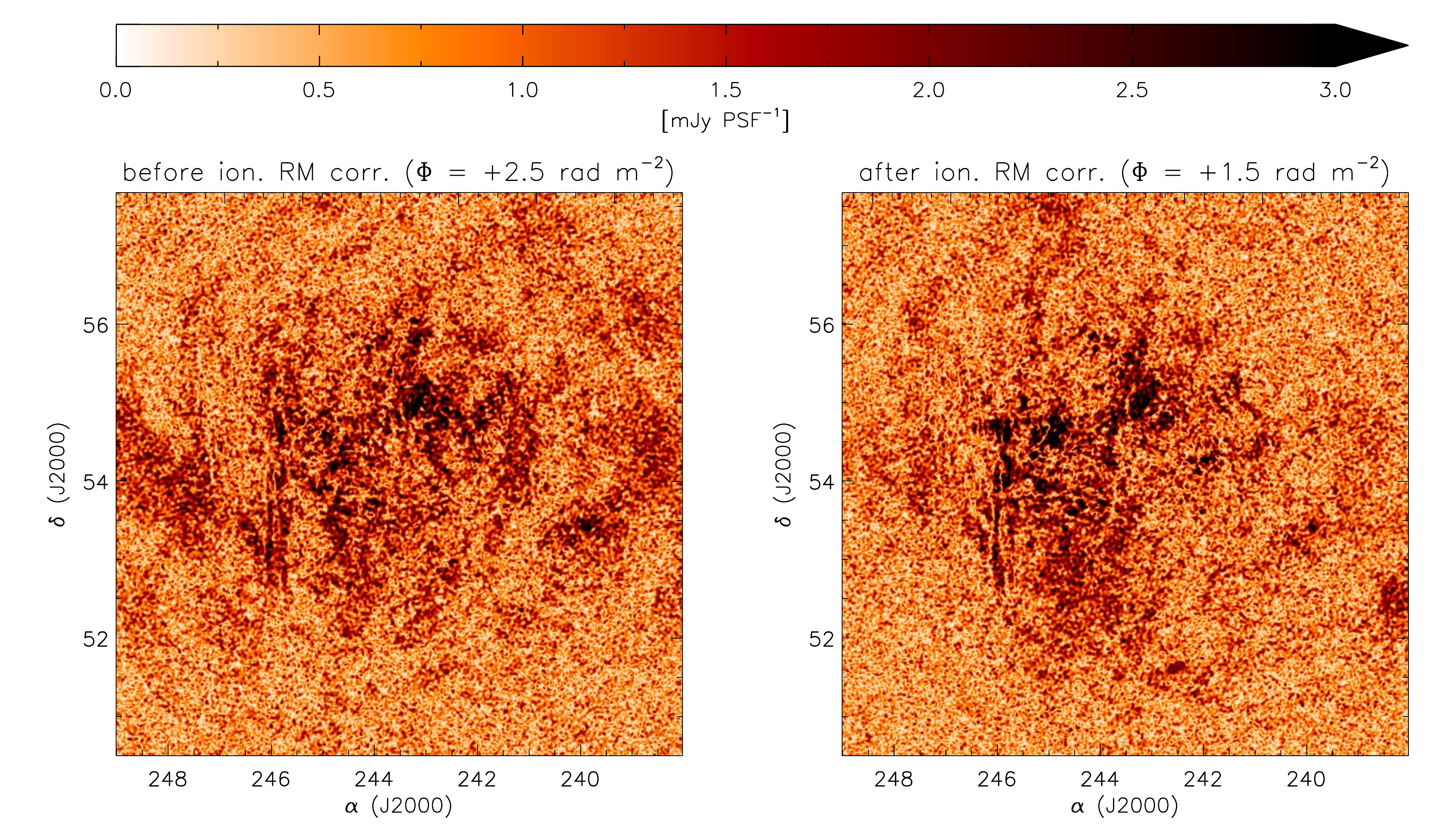}
\centering \includegraphics[width=.86\textwidth]{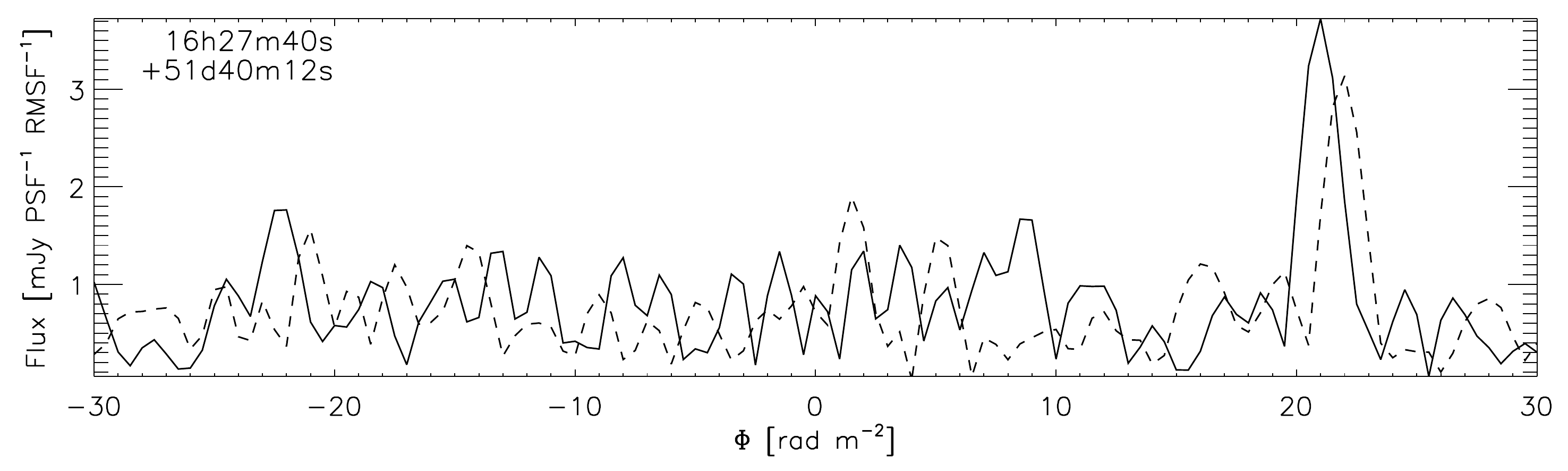}
\caption{Effect of the ionospheric RM variation. Bottom plot shows the Faraday spectrum centred at the SW lobe of the giant radio galaxy J162740+514012, before (dashed line) and after (solid line) applying the ionospheric RM correction. As expected, there is (i) a shift in Faraday spectrum that corresponds to an average of the ionospheric rotation measure; and (ii) an increase in the peak flux by 20\%. The ionospheric RM variation also has an effect on the diffuse polarized emission as shown in the two images in the upper part of the figure. Note that these images are given at Faraday depths separated by an average of the ionospheric rotation measure to correct for a shift due to Faraday rotation in the ionosphere.}
\label{fig:ionRMcorr}
\end{figure*}

\section{Results}
\subsection{Widefield image in total intensity}

\begin{figure*}[!phtb]
\centering \includegraphics[width=.33\textwidth]{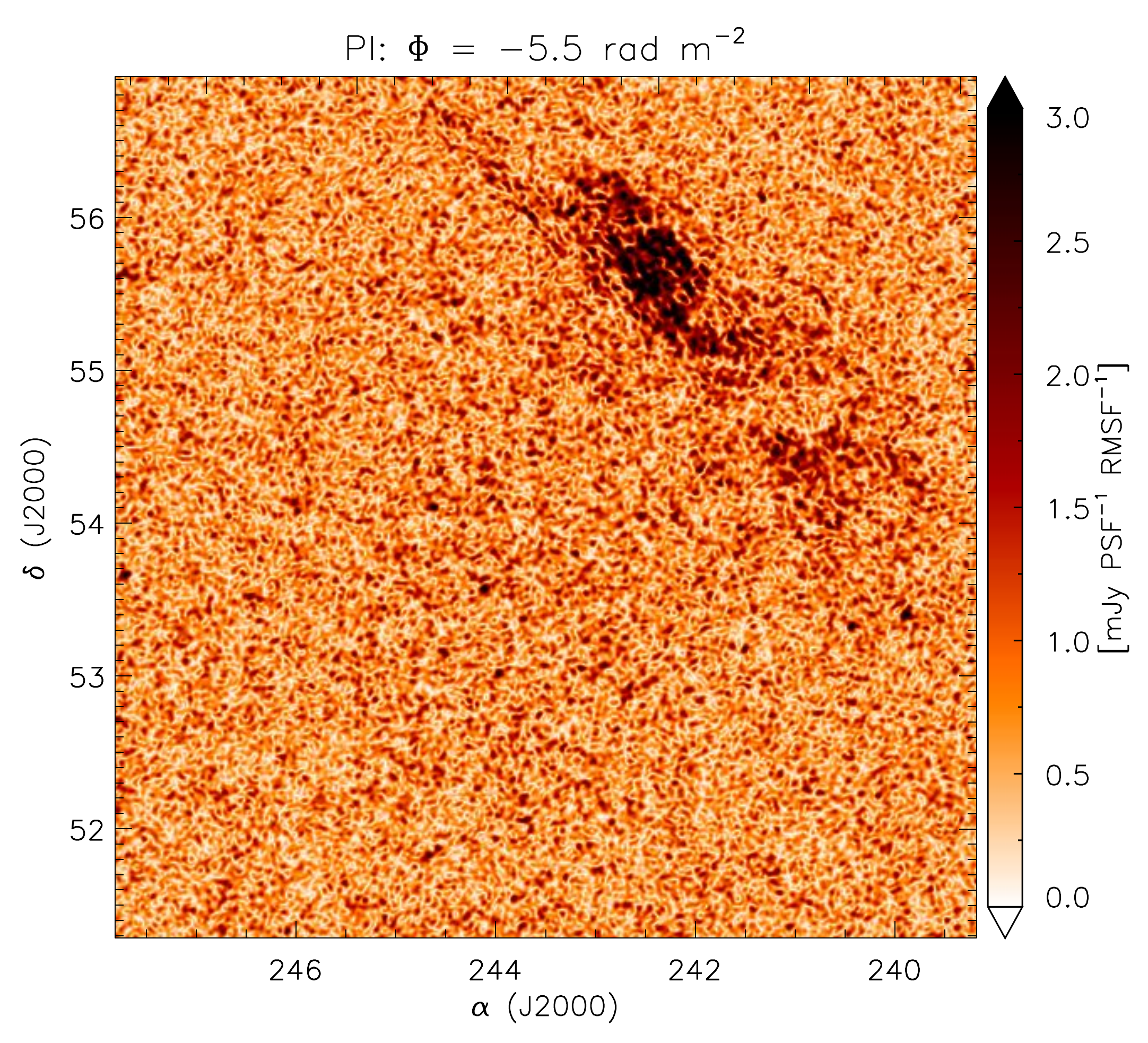}
\centering \includegraphics[width=.33\textwidth]{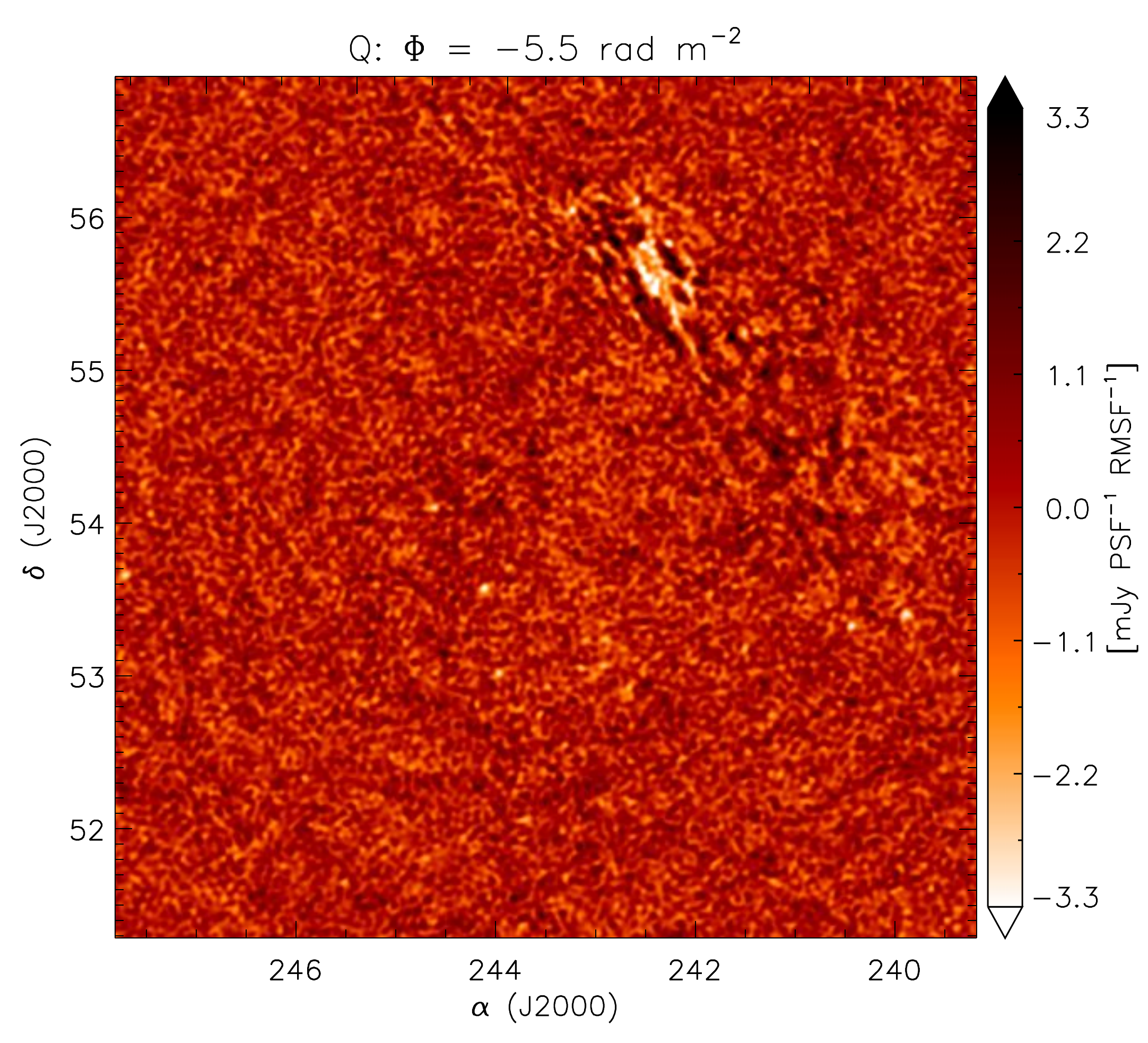}
\centering \includegraphics[width=.33\textwidth]{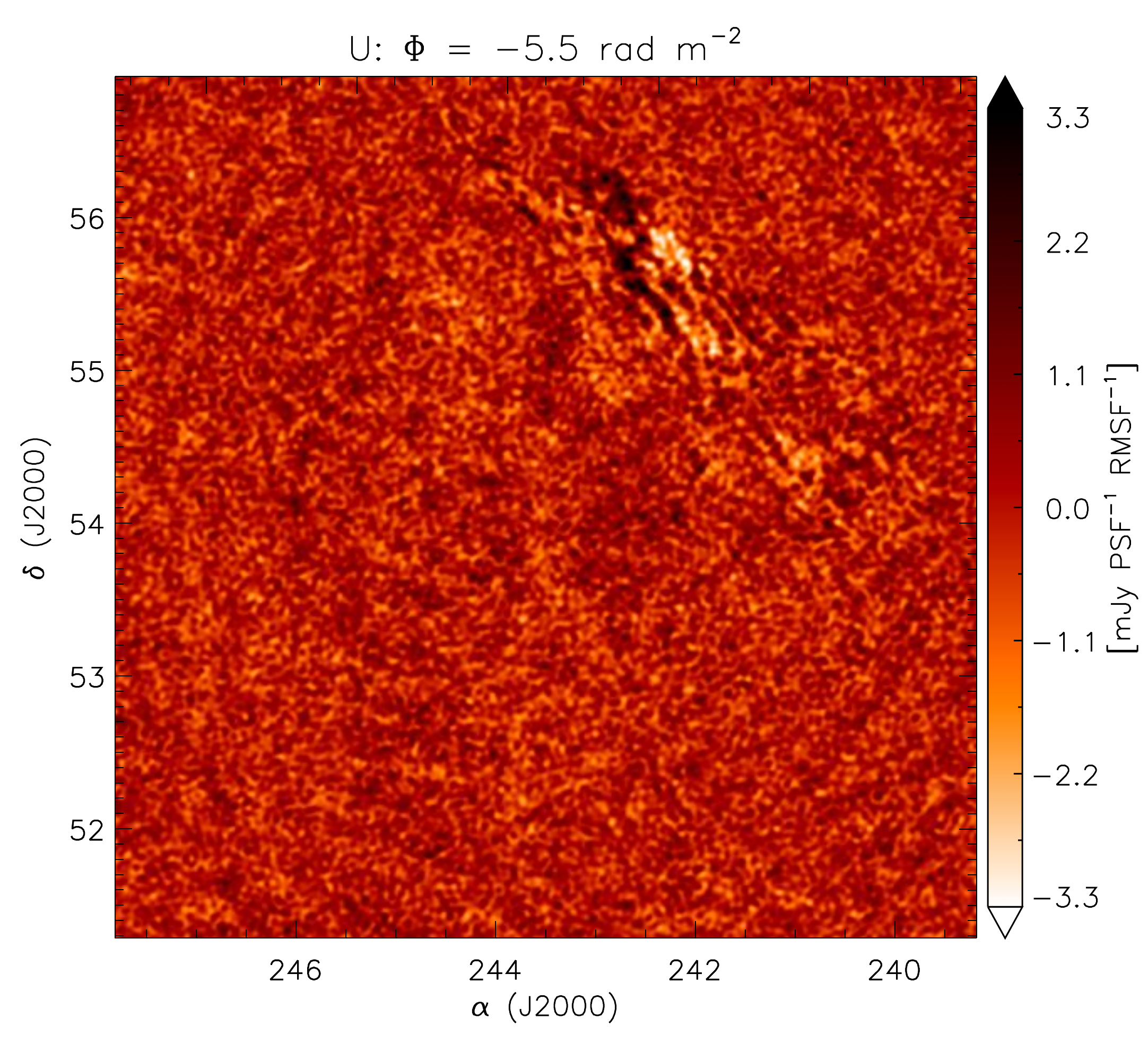}
\centering \includegraphics[width=.33\textwidth]{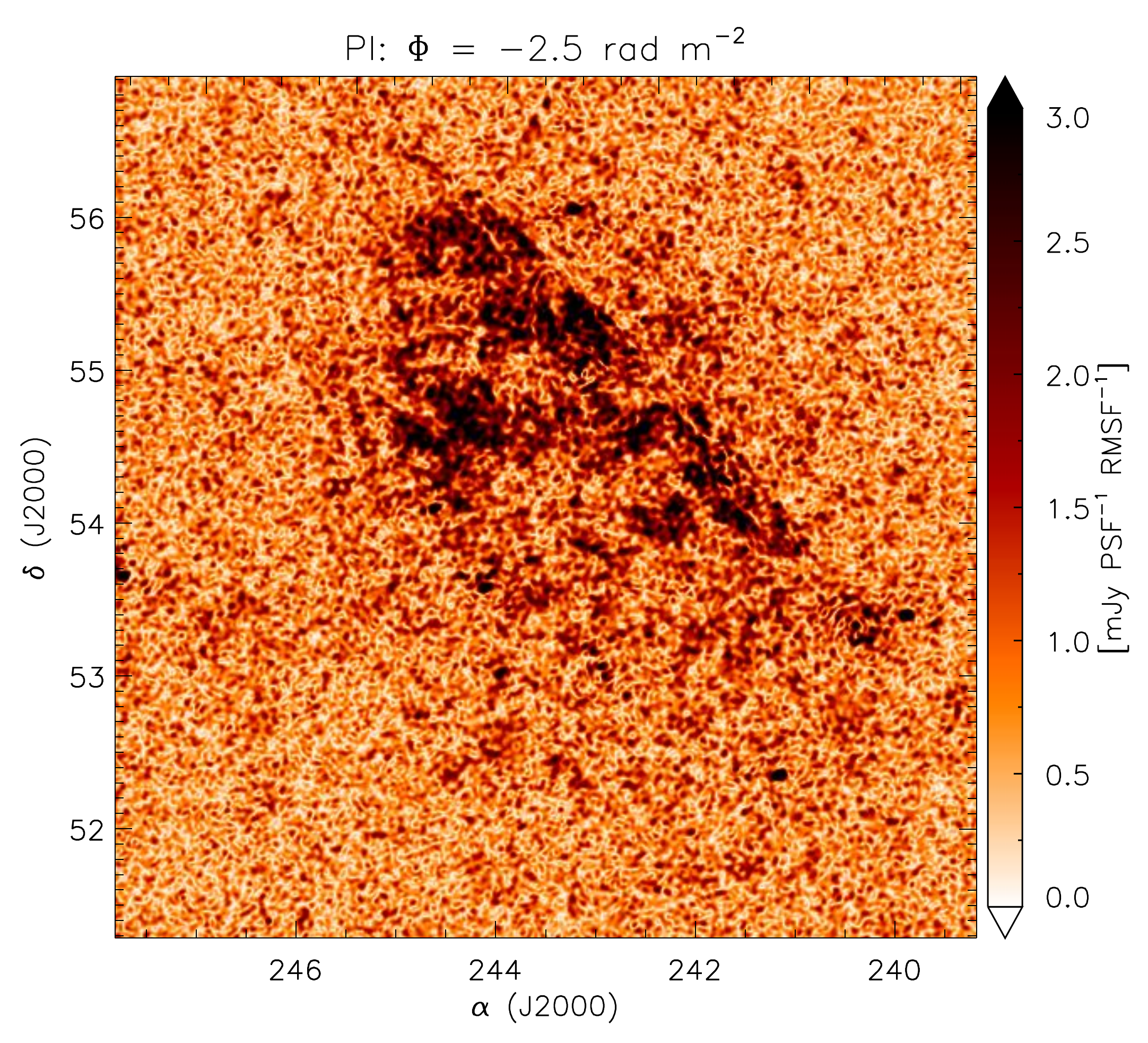}
\centering \includegraphics[width=.33\textwidth]{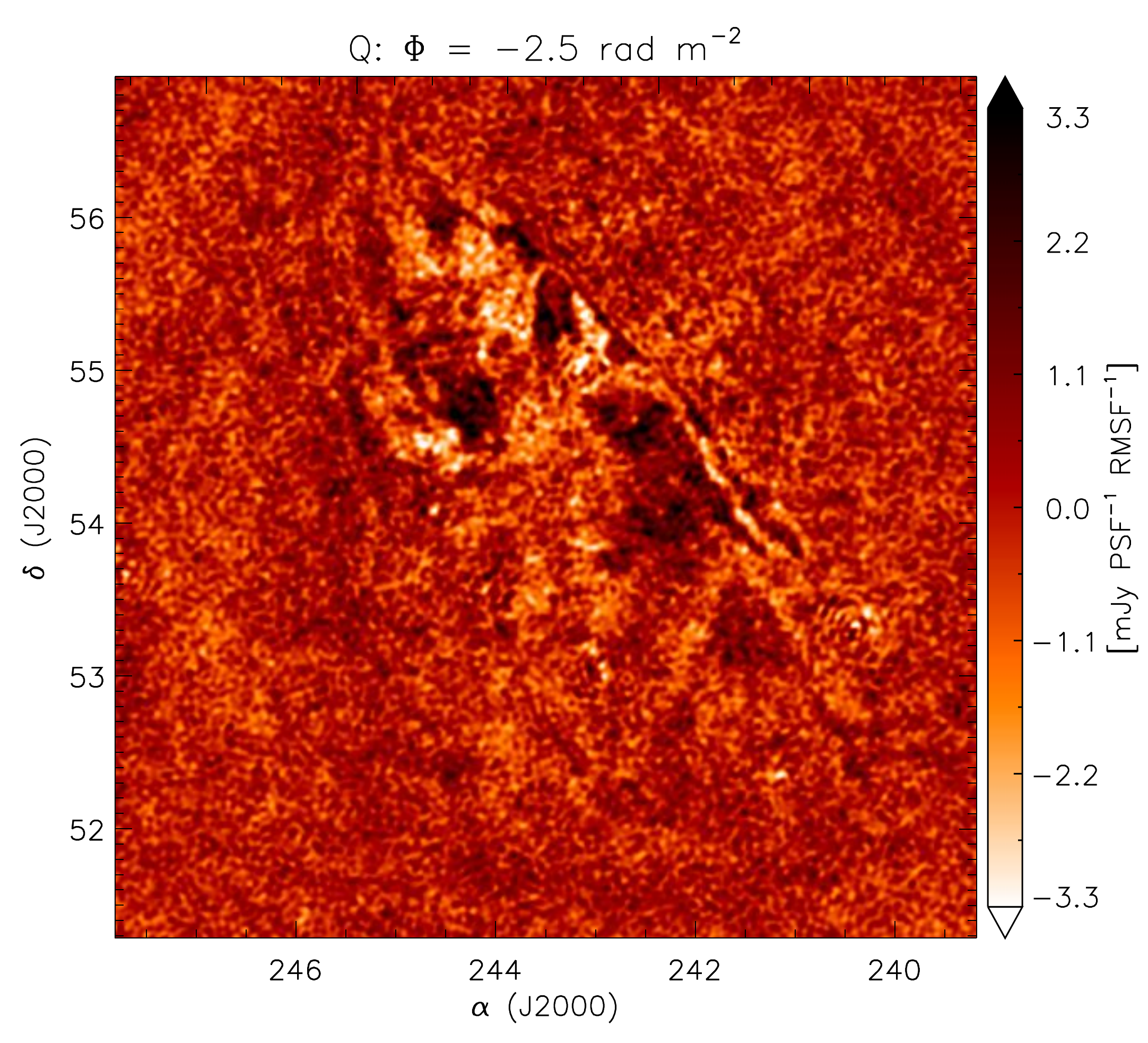}
\centering \includegraphics[width=.33\textwidth]{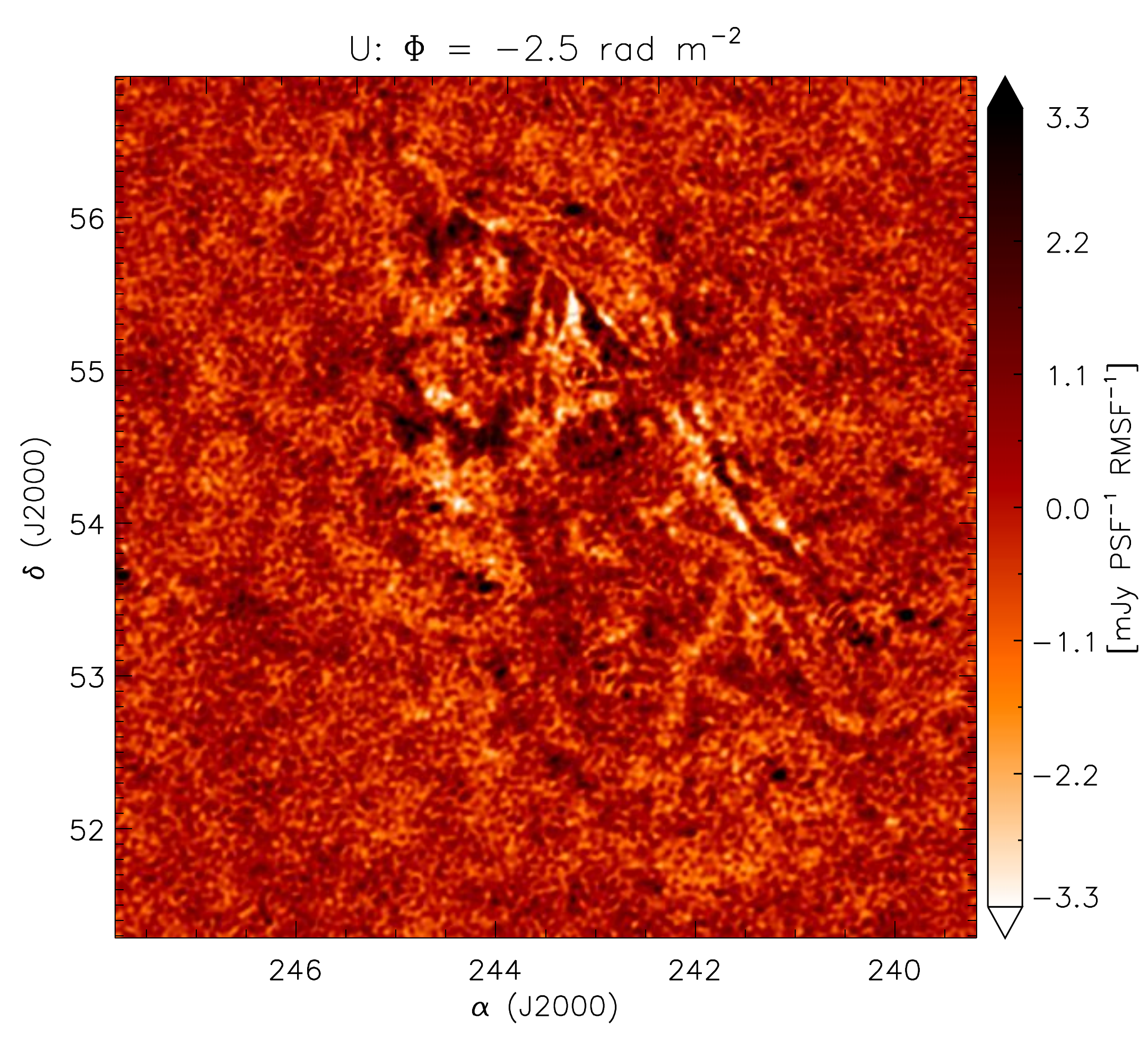}
\centering \includegraphics[width=.33\textwidth]{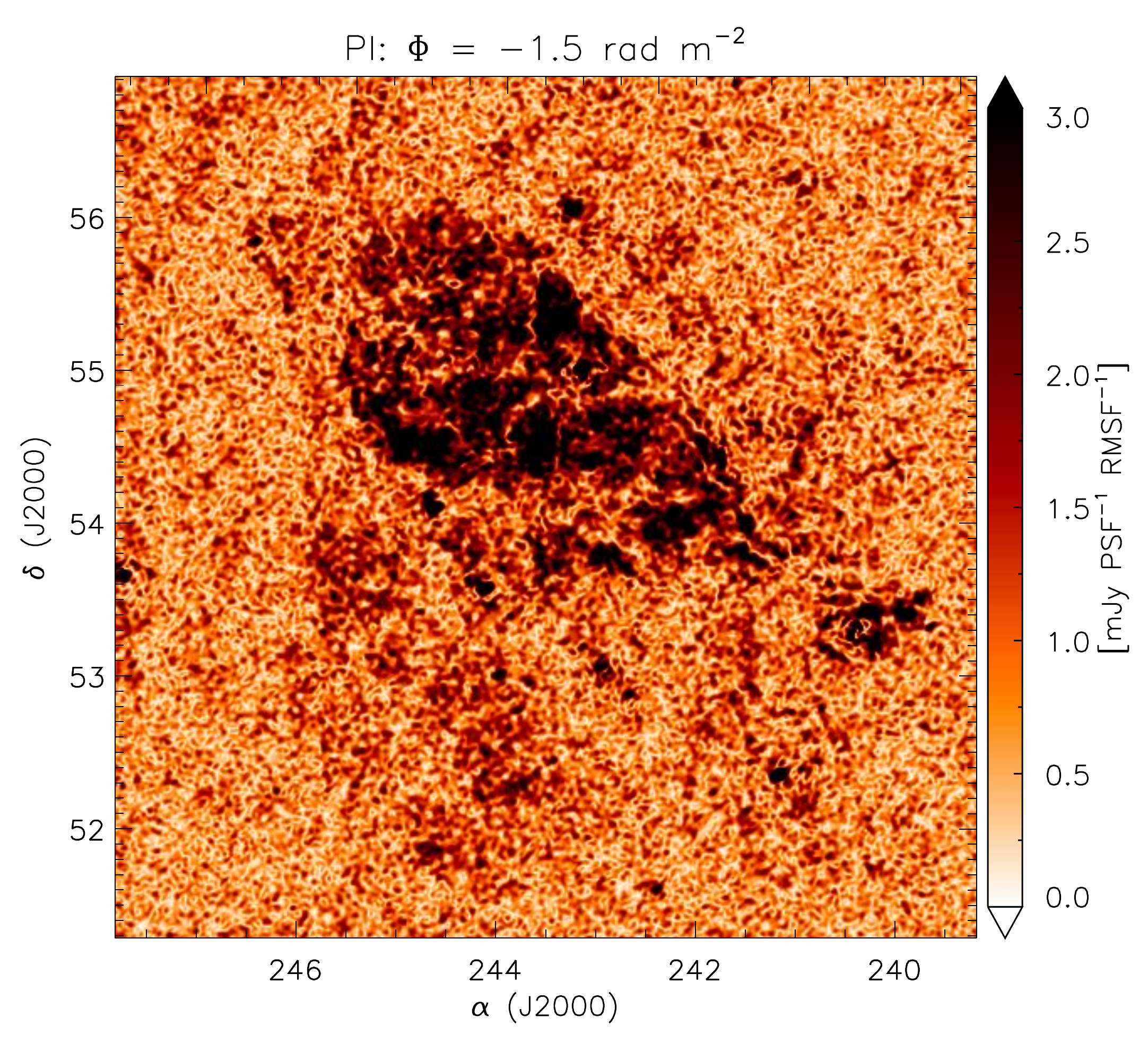}
\centering \includegraphics[width=.33\textwidth]{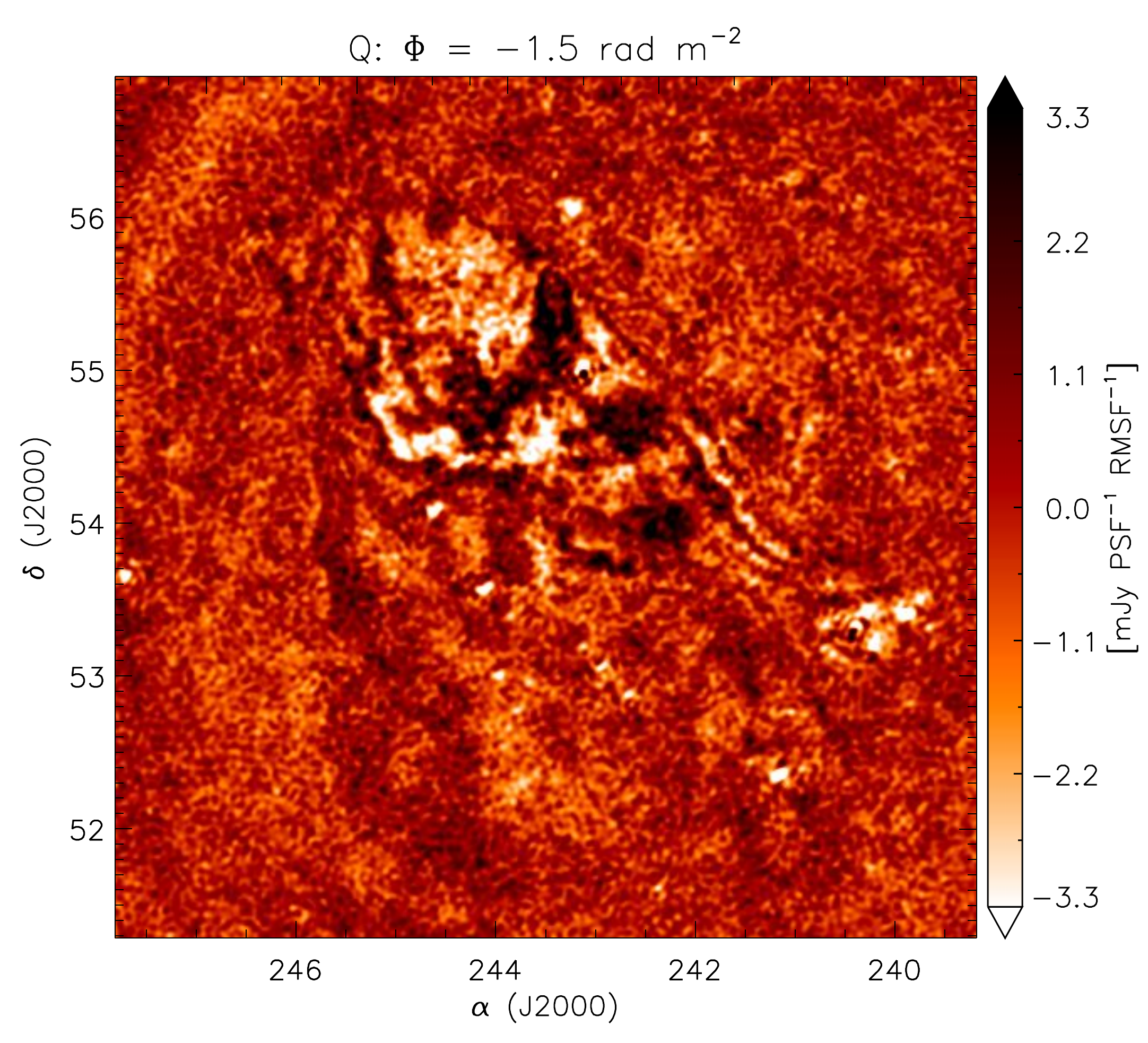}
\centering \includegraphics[width=.33\textwidth]{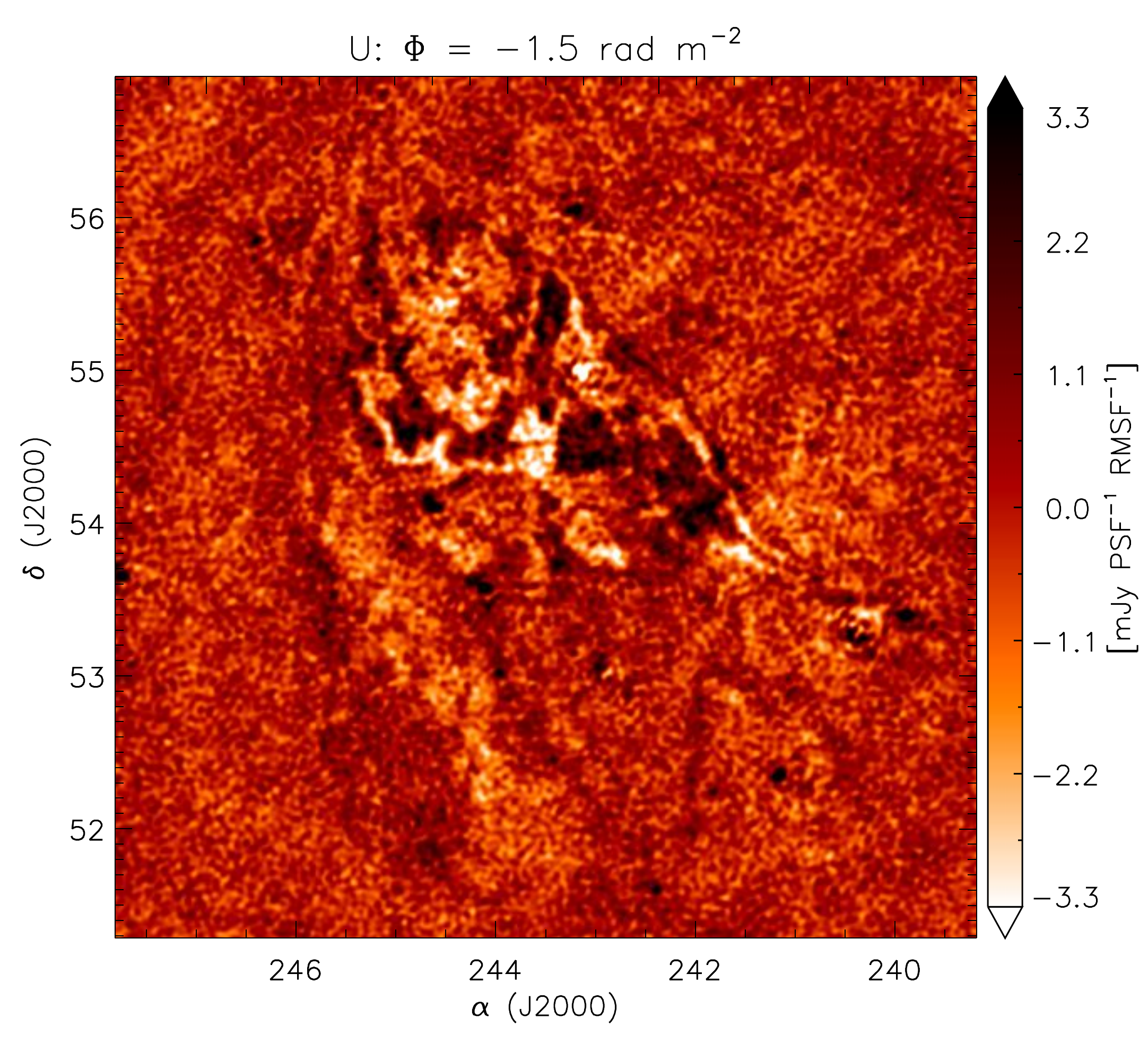}
\centering \includegraphics[width=.33\textwidth]{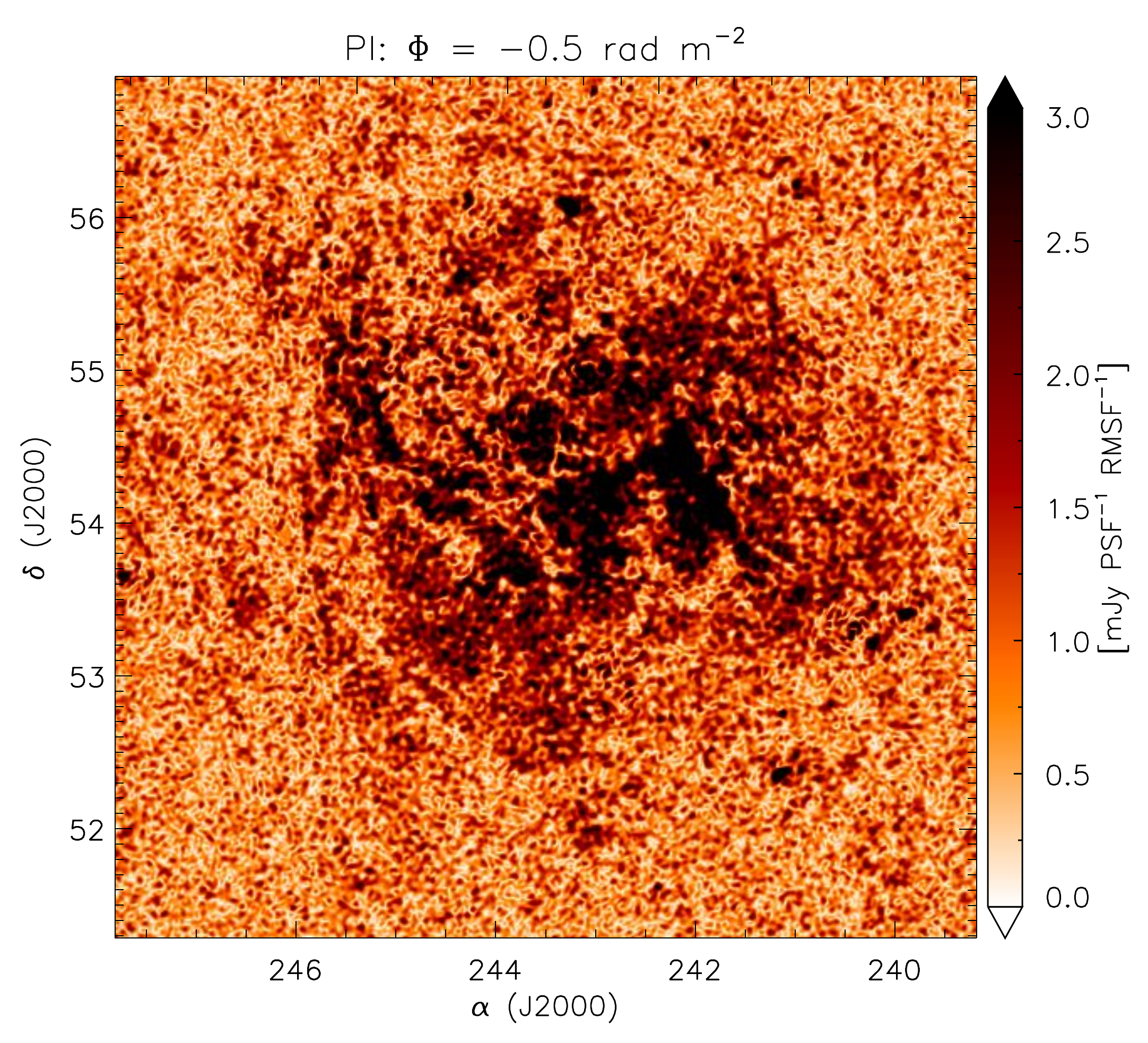}
\centering \includegraphics[width=.33\textwidth]{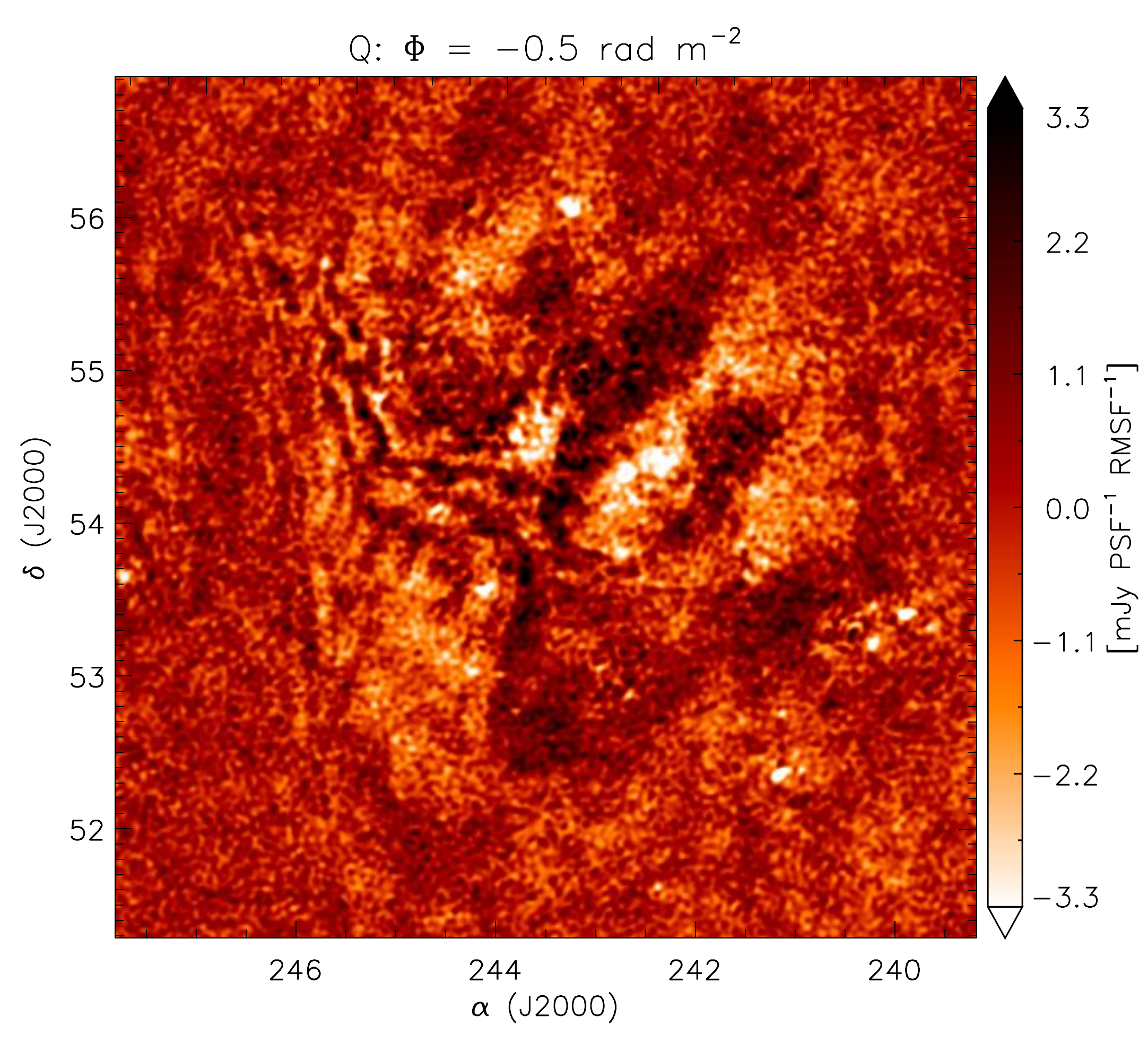}
\centering \includegraphics[width=.33\textwidth]{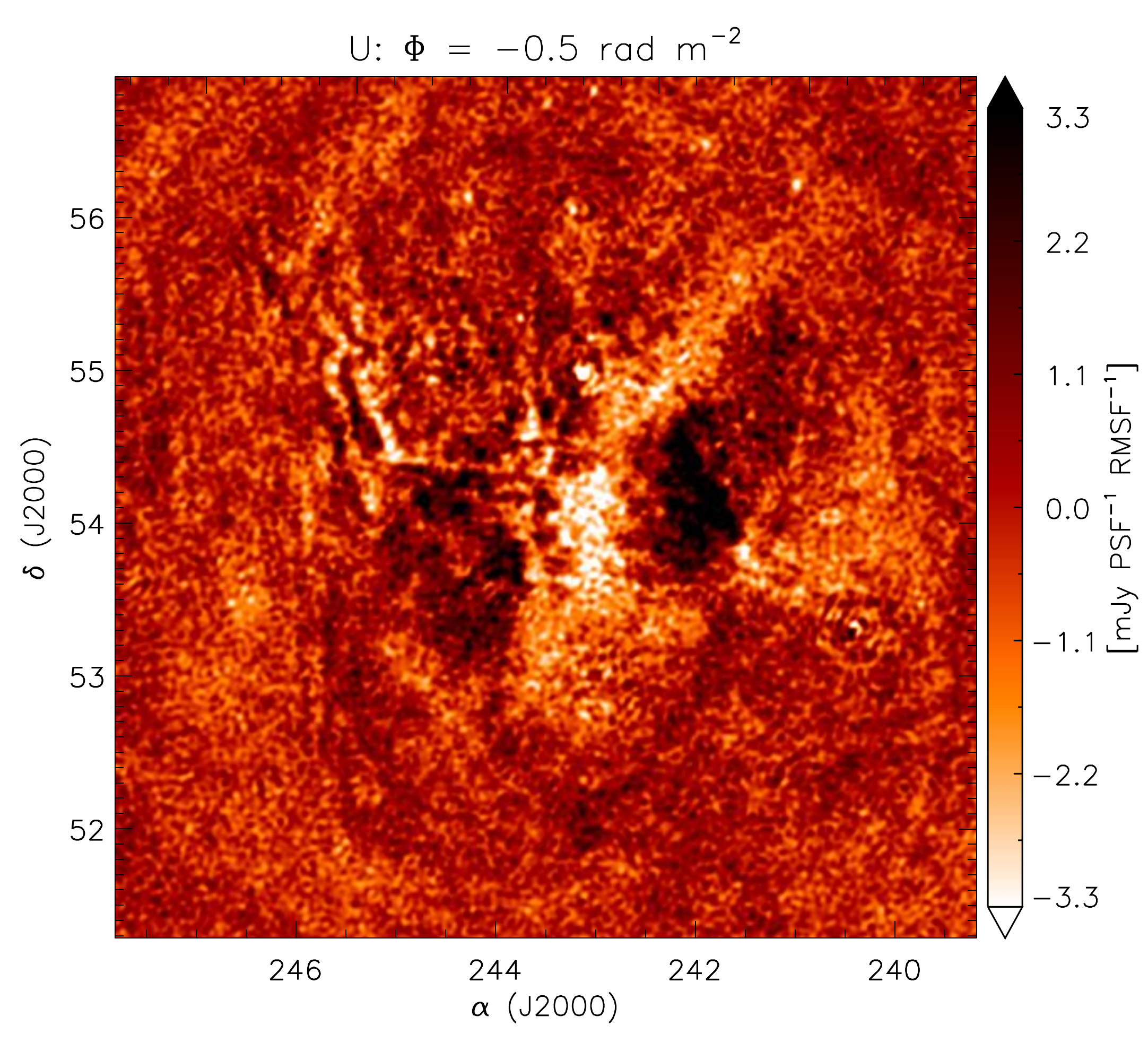}
\caption{Widefield images of the ELAIS-N1 region in polarized intensity (PI), Stokes Q, and Stokes U given at Faraday depths
of -5.5, -2.5, -1.5, -0.5, +0.5, +1.5~${\rm rad~m^{-2}}$. Images are $5.7^\circ\times5.7^\circ$ in size with a PSF of $3.4'\times3.1'$. 
The noise level is $0.3~{\rm mJy~PSF^{-1}~RMSF^{-1}}$ in polarized intensity and $0.5~{\rm mJy~PSF^{-1}~RMSF^{-1}}$ in Stokes Q,U. 
The images in polarized intensity have not been corrected for polarization noise bias.}
\label{fig:PI}
\end{figure*}

\addtocounter{figure}{-1}
\begin{figure*}{}
\centering \includegraphics[width=.33\textwidth]{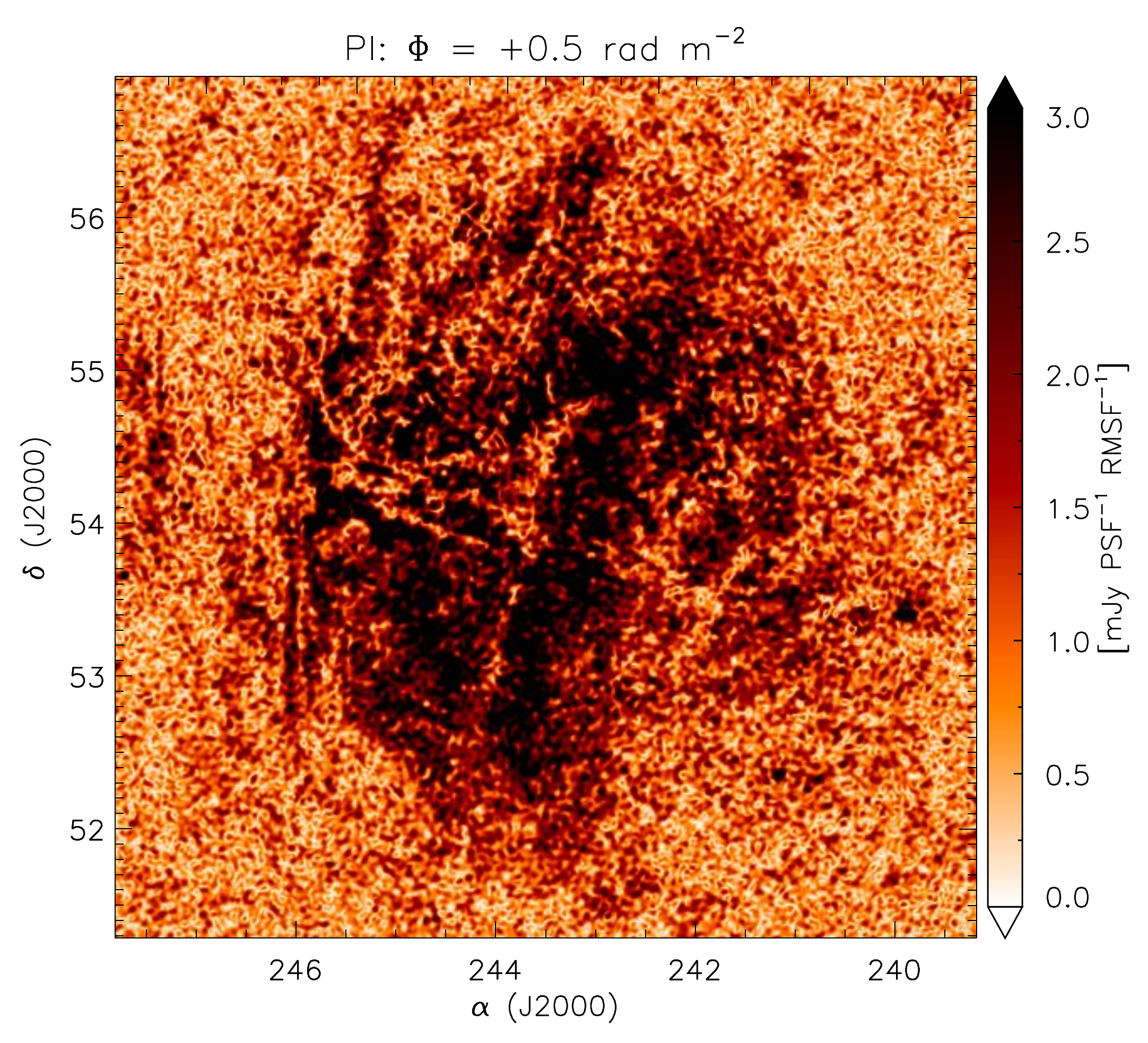}
\centering \includegraphics[width=.33\textwidth]{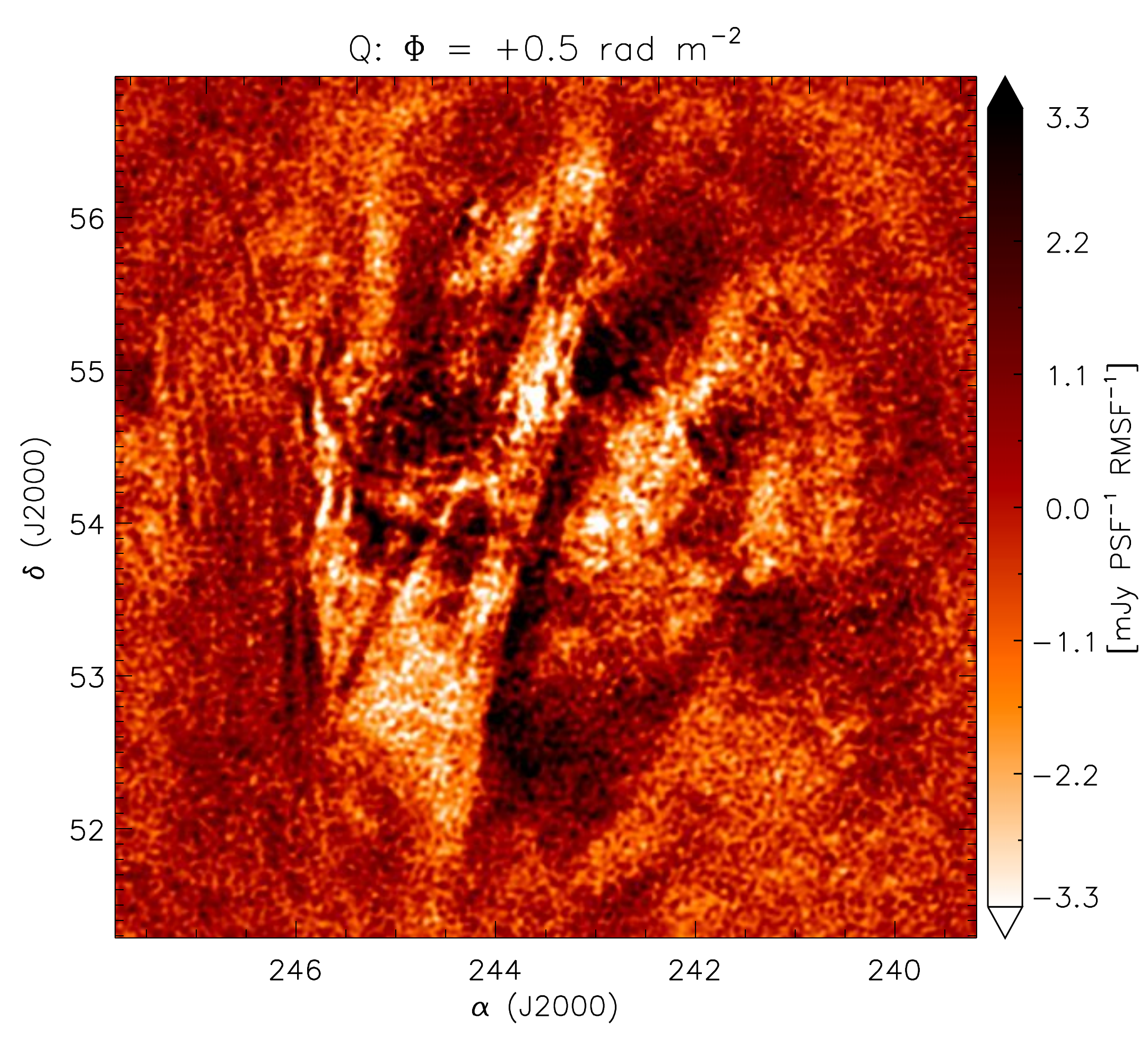}
\centering \includegraphics[width=.33\textwidth]{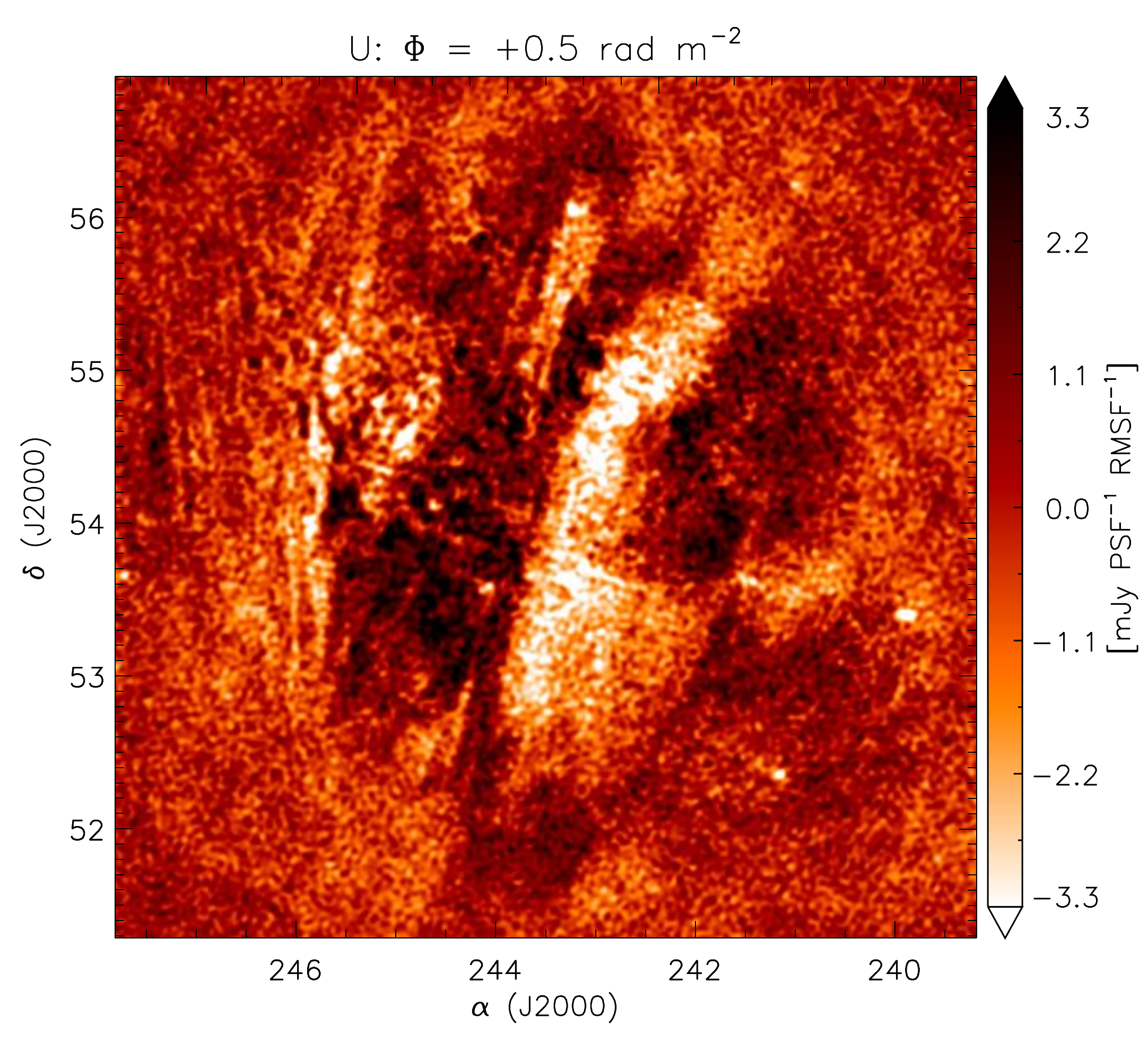}
\centering \includegraphics[width=.33\textwidth]{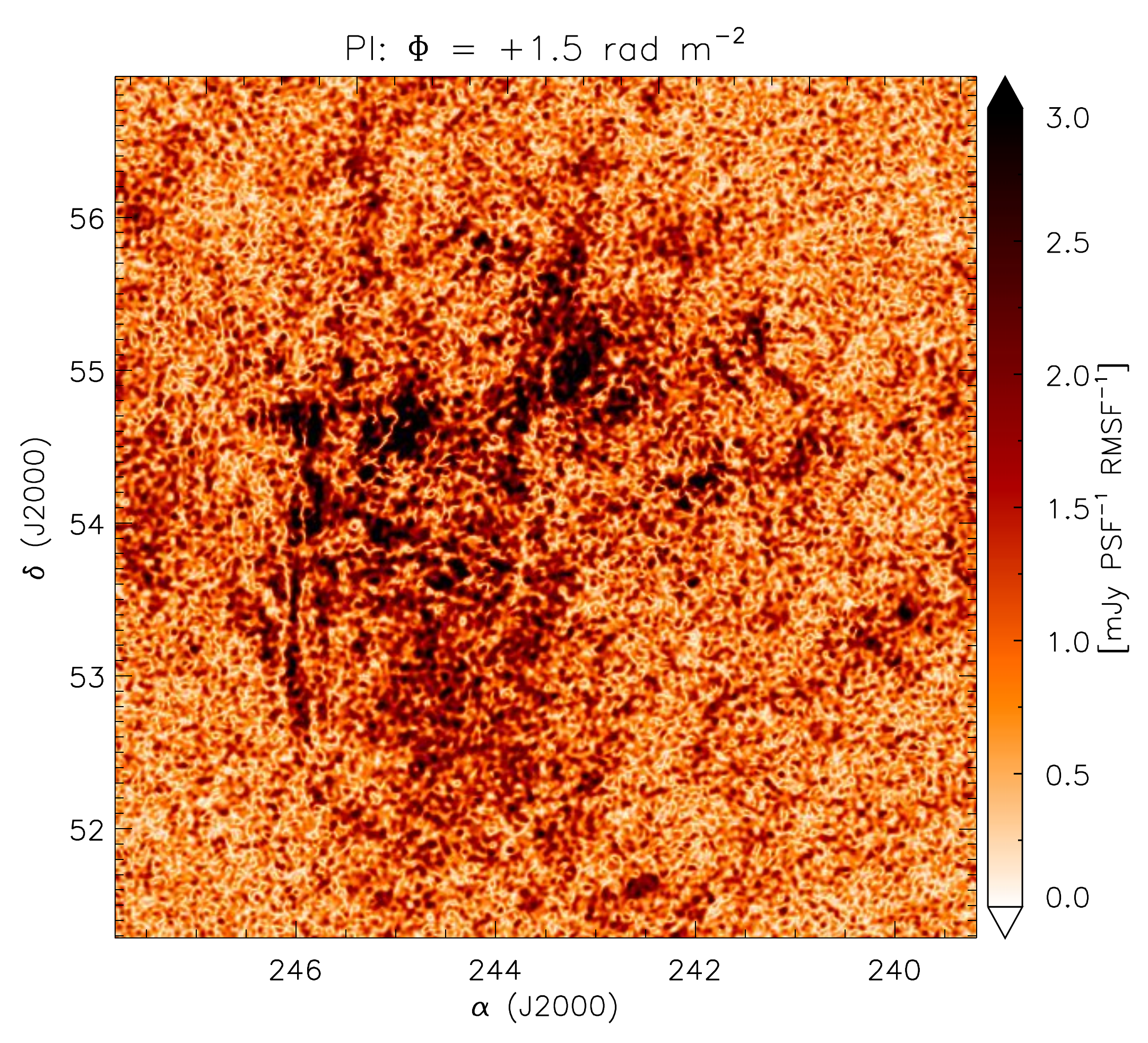}
\centering \includegraphics[width=.33\textwidth]{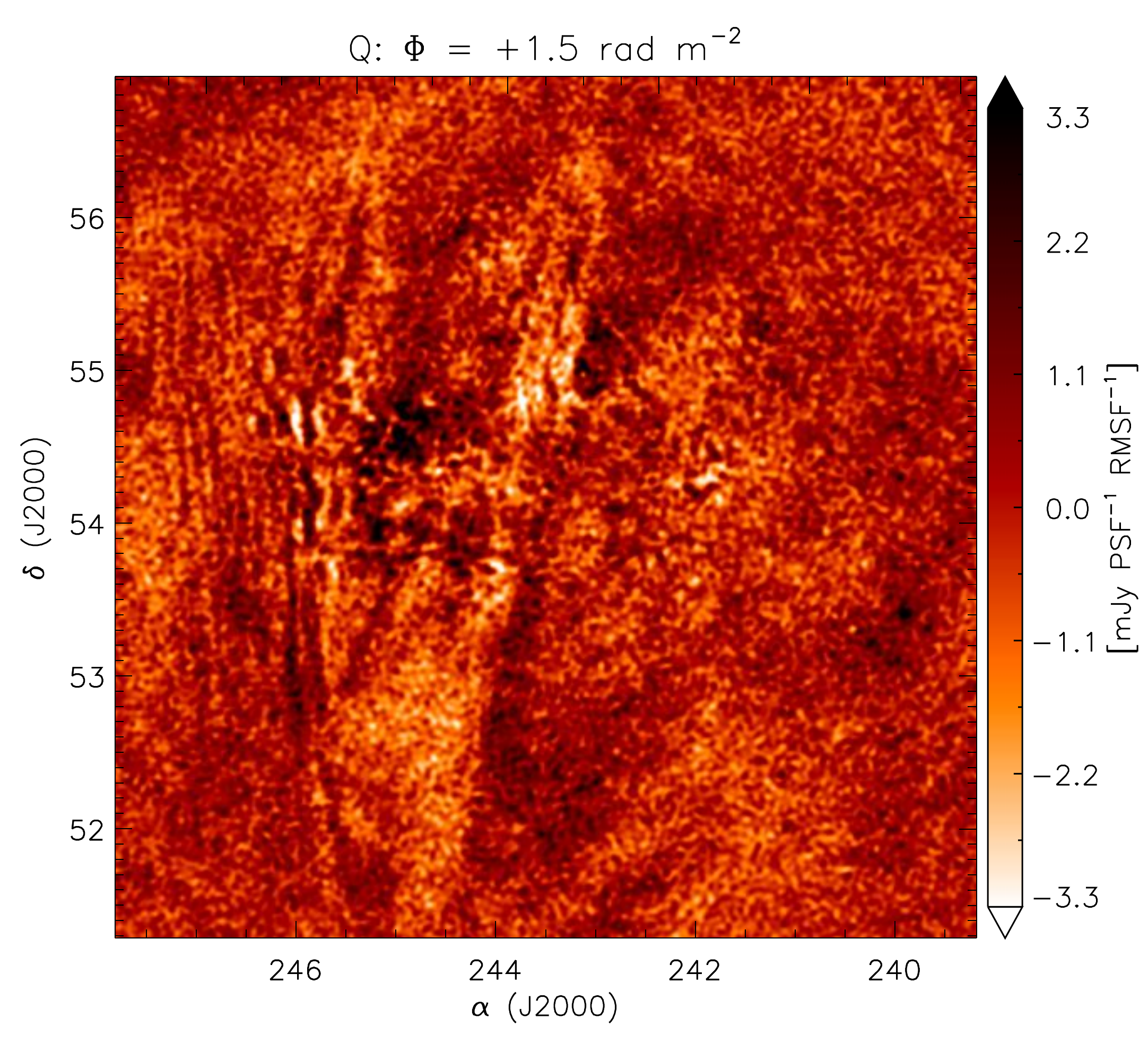}
\centering \includegraphics[width=.33\textwidth]{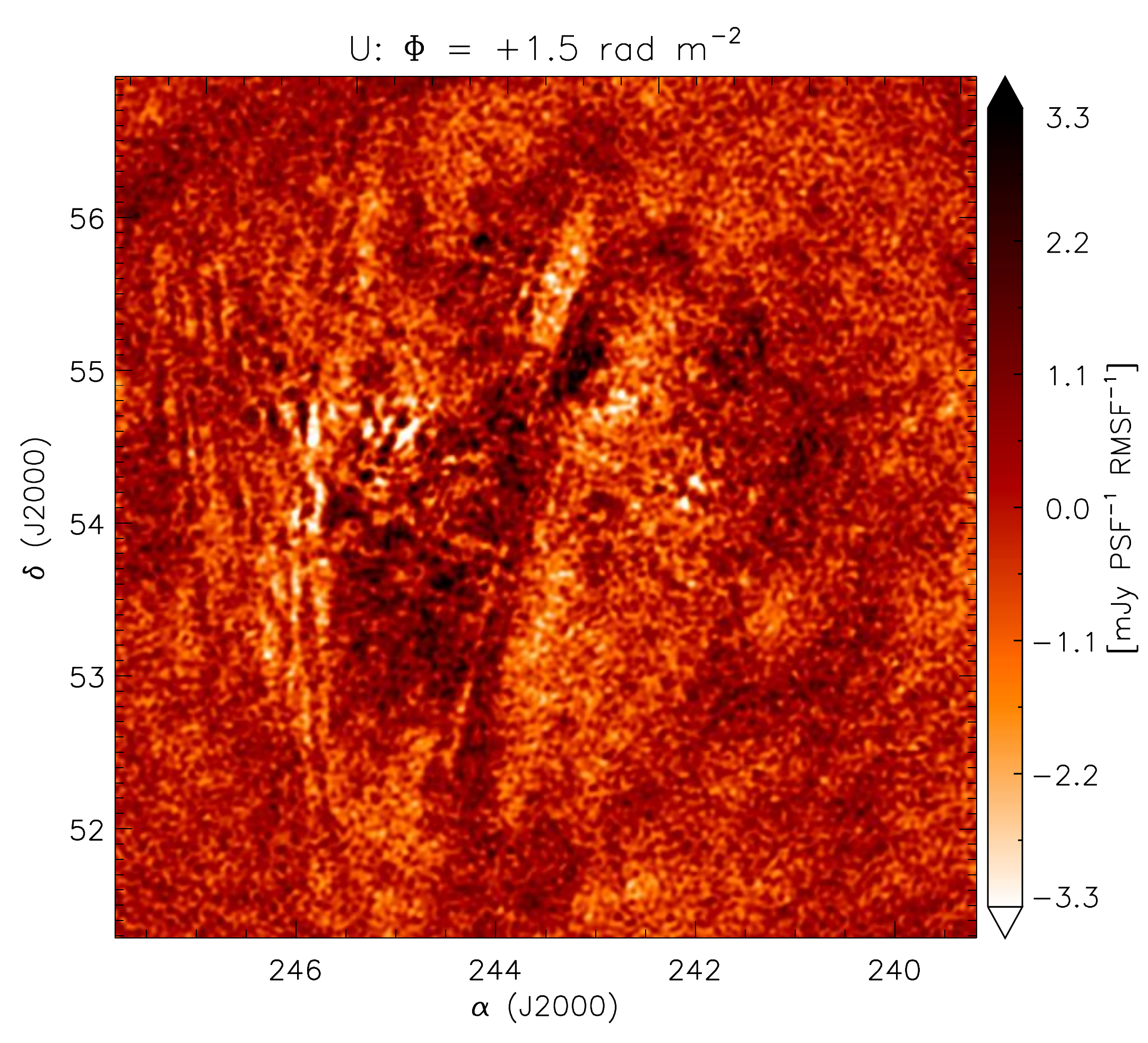}
\caption{Continued.}
\label{fig:PI2}
\end{figure*}

The frequency-averaged Stokes I widefield image of the ELAIS-N1 region is presented in Fig.~\ref{fig:stokesI}. 
The image is obtained after calibration as described in Sect.~\ref{sec:cal} and it is $6.6^\circ\times6.6^\circ$ in size with a PSF of $16.0''\times8.8''$. The noise level is $1.0~{\rm mJy~PSF^{-1}}$. There is no indication of diffuse emission in total intensity. A circle marks a region around the giant radio galaxy J162740+514012 \citep[see][]{schoenmakers01}, which contains two lobes that were found to be highly polarized at 325 MHz. The source is found to be polarized  in the LOFAR band as well, although at much lower levels. 
The RMs of the two lobes have been determined at both frequency regimes. For each lobe, the two determinations at two different frequency regimes agree to within the measurement accuracy. Small calibration errors due to  variation of the station beams and rapid ionospheric phase fluctuations  are still visible around some bright sources. These errors can be suppressed by direction dependent calibration using \texttt{SageCal}. For the purpose of this paper we are mainly interested in polarization, and we will leave the analysis of the total intensity emission for future work. 

\begin{figure}[tb]
\centering \includegraphics[width=.5\textwidth]{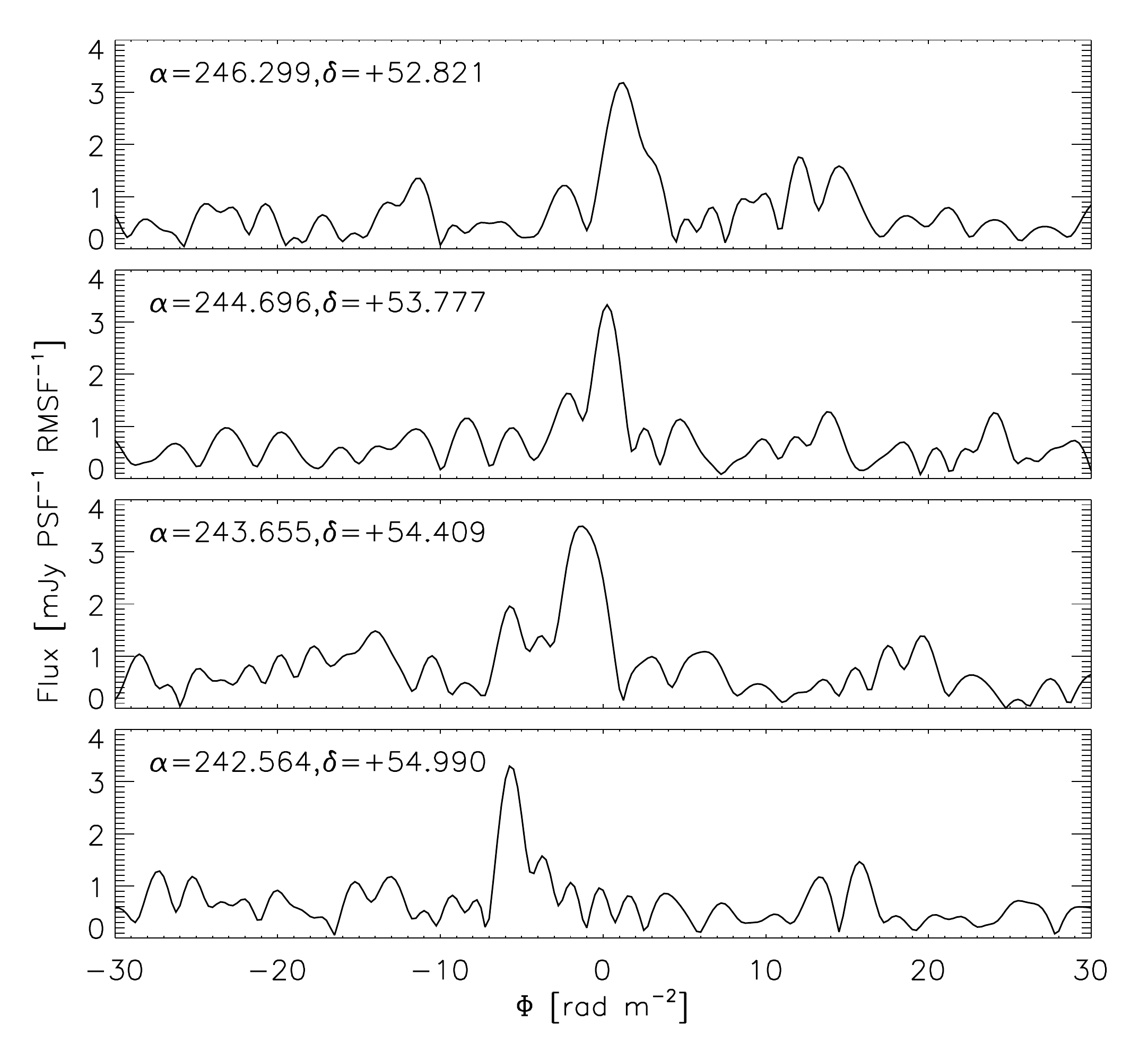}
\caption{Faraday depth spectra of a few interesting lines of sight through the RM cube of polarized intensity.
The typical width of the detected structures is a few ${\rm rad~m^{-2}}$.}
\label{fig:Fspec}
\end{figure}

\subsection{Correcting for Faraday rotation in the Earth's ionosphere}\label{sec:ion}
An electromagnetic wavefront passing through an ionized medium with a variable index of refraction experiences time delays in different parts of the wavefront. These delays are proportional to the total electron content (TEC)\footnote{Total electron content (TEC) is the integrated electron density along the line of sight through Earth's ionosphere, with units of electrons per square meter ($10^{16}~{\rm electrons/m^2} = 1~{\rm TEC~unit}$). } and are inversely proportional to the square of the observing frequency  \citep[e.g.][]{thompson07}. Indeed, at LOFAR frequencies the Earth's ionosphere is the dominant source of phase errors. A linear spatial ionospheric TEC gradient causes a position shift of the source. Higher-order variations in the index of refraction cause a more serious distortion producing defocusing and even scintillations in extreme cases. 

In polarimetric studies there is an additional ionospheric effect that one needs to correct for. Faraday rotation in the Earth's ionosphere changes the polarization angle of  the incoming polarized emission. This happens on a timescale that is much smaller than the total integration time of an observation. As a result of this, the observed polarized emission will  be shifted in Faraday depth space and be partially decorrelated. The average shift is proportional to the ionospheric RM averaged over the observing time. If the variation during the synthesis time is longer than $\approx 1$~rad the emission will be depolarized and the dynamic range in the image will be reduced. Thus, one needs to estimate the amount of ionospheric RM as a function of time and then ``derotate'' the observed polarization angle by that amount. Large TEC-gradients  across the array can cause differential Faraday rotation, making unpolarized sources appear circularly polarized. These were not present in the current data. To estimate the ionospheric RM during our observation, we use the Global Ionospheric Maps (GIMs) of the vertical total electron content  (VTEC) and the Earth's magnetic field model \citep[International Geomagnetic Reference Field; ][]{maus05} from the \texttt{casacore}\footnote{http://casacore.googlecode.com} library.  The GIMs are generated by the Royal Observatory of Belgium using near real time data taken every $15~{\rm min}$ by more than 100 Global Positioning System (GPS) sites spread across Europe \citep[][http://gnss.be]{bergeot09}. The resolution of these maps is $0.5^\circ\times0.5^\circ$ in longitude and latitude.

The estimated ionospheric RM values for the ELAIS-N1 observation are plotted in Fig.~\ref{fig:ionRM}. The difference between the RM values across the LOFAR array  (presented with different lines in Fig.~\ref{fig:ionRM}) is very small for this particular observation. 
The degree of ionization in the ionosphere depends essentially on the amount of radiation received from the Sun. Hence, the RM values are higher at the beginning of the observation ($\sim2.5{\rm h}$ before sunset) than at the end of it ($\sim4.5{\rm h}$ after sunset). During our observation the ionosphere shows typical behaviour and the estimated ionospheric RM corrections are not unusual.

\begin{figure*}[tb]
\centering \includegraphics[width=.45\textwidth]{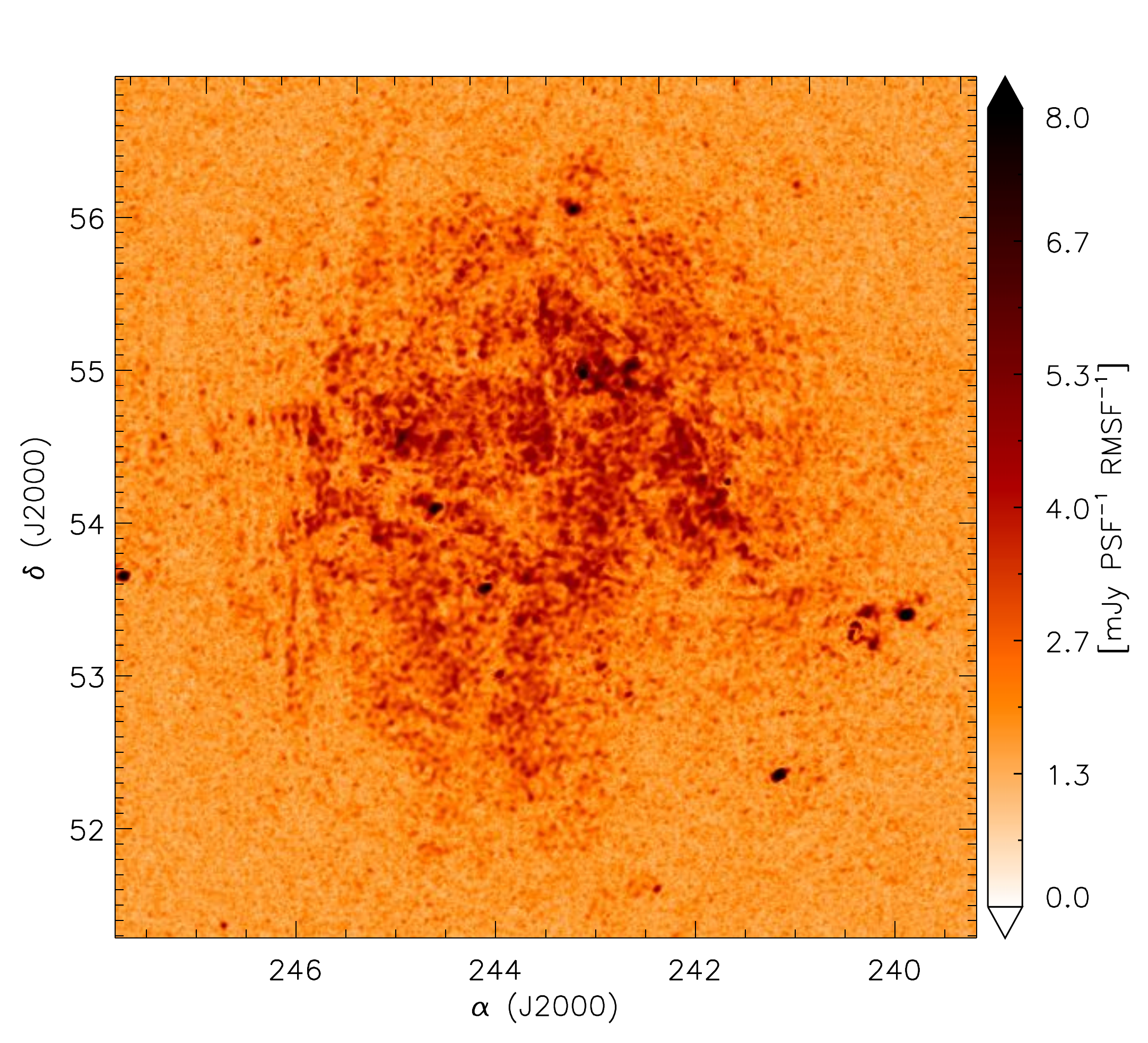}
\centering \includegraphics[width=.45\textwidth]{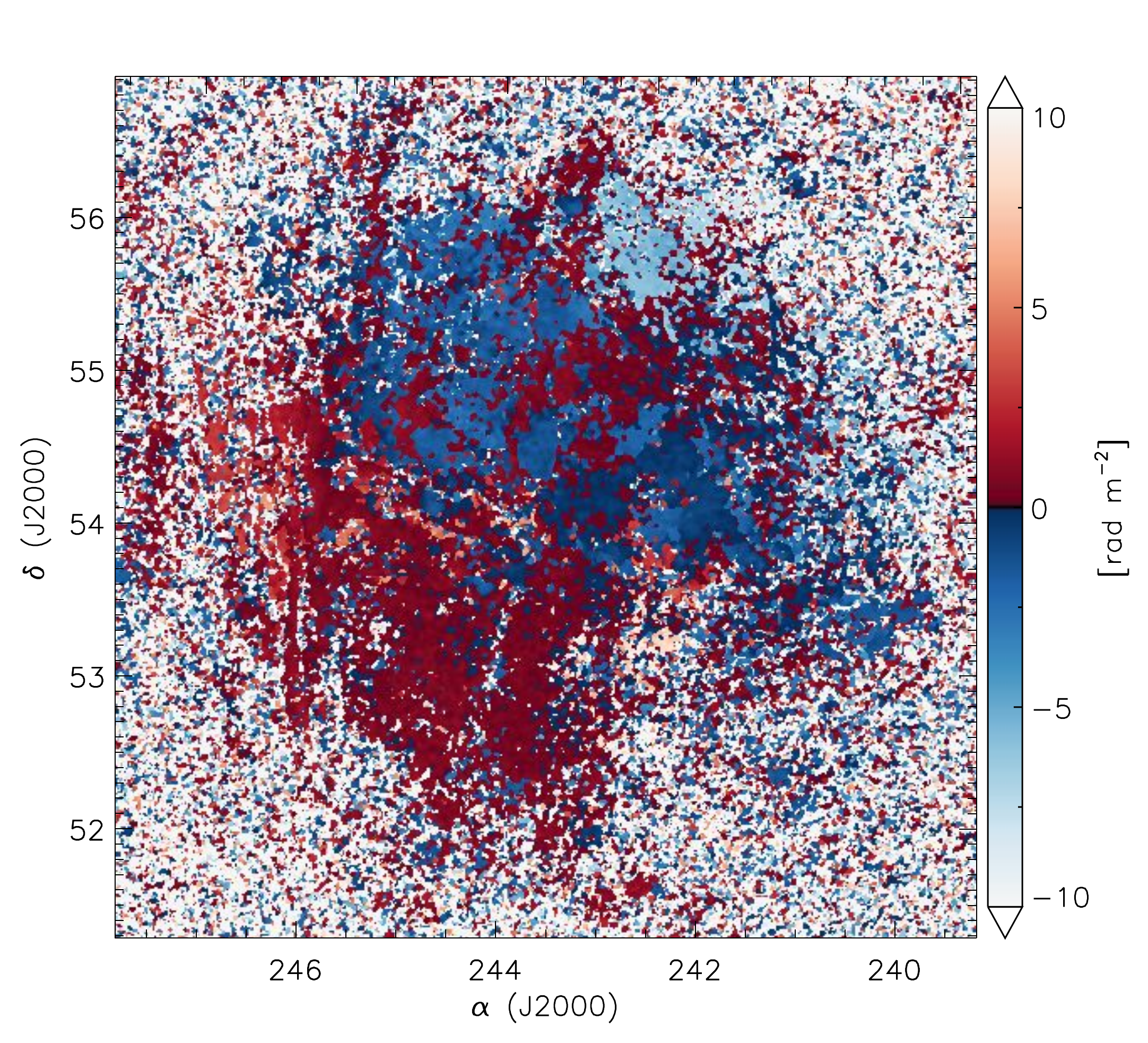}
\caption{Image on the left shows the distribution of peak intensities in the Faraday depth spectra at every spatial pixel.  Image on the right gives the corresponding RM value for each pixel}
\label{fig:RMmax}
\end{figure*}

We have corrected our data for the time-variable ionospheric RM using \texttt{RMWriter} code written by M. Mevius. Figure~\ref{fig:ionRMcorr} shows the Faraday spectrum centred at the SW (Southwest) lobe of the giant radio galaxy J162740+514012. This lobe is polarized and  its emission appears around $\Phi=+22~{\rm rad~m^{-2}}$ before applying the ionospheric RM correction (dashed line in Figure~\ref{fig:ionRMcorr}). After we apply the ionospheric RM correction (solid line in Figure~\ref{fig:ionRMcorr}), its emission is centred around $\Phi=+21~{\rm rad~m^{-2}}$. A shift of $\left|\Delta\Phi\right|=1~{\rm rad~m^{-2}}$ corresponds to the average of the ionospheric rotation measure $\left<{\rm RM_{ion}}\right>=1.2~{\rm rad~m^{-2}}$. There is also an increase in the peak flux by 20\%. 

Correcting the data for the time-dependent ionospheric RM variation also has an effect on the images of the diffuse polarized emission. An example is shown in Figure~\ref{fig:ionRMcorr}. For the predicted RM variation we would expect this effect to be about 15-20\%.  Indeed, the diffuse emission is brighter, the morphological features are sharper and the edge of the station primary beam is more clearly visible in the corrected image, as compared to the uncorrected image.  

\subsection{Diffuse polarized  emission} 
A series of widefield images of the ELAIS-N1 region in both polarized intensity and Stokes Q,U are presented in Fig.~\ref{fig:PI}. 
Given the observed rather uniform levels of the polarized intensity the polarity of Stokes Q,U is a good 
indicator of the spatial variations of the plane of polarization.
The images are $5.7^\circ\times5.7^\circ$ in size with a PSF of $3.4'\times3.1'$ and the noise level
is $0.3~{\rm mJy~PSF^{-1}~RMSF^{-1}}$ in polarized intensity and $0.5~{\rm mJy~PSF^{-1}~RMSF^{-1}}$ in Stokes Q,U. 
Images are given at Faraday depths of -5.5, -2.5, -1.5, -0.5, +0.5, +1.5~${\rm rad~m^{-2}}$ to emphasise
the various detected structures in linear polarization. Note that all images are produced using the RM cubes
corrected for Faraday rotation in the ionosphere.

We detect faint polarized emission ($\lesssim 1~{\rm mJy~PSF^{-1}~RMSF^{-1}}$) over  a range of 
Faraday depths from $-10$ to $+13~{\rm rad~m^{-2}}$. The brightest and most prominent features 
are detected in a smaller range of Faraday depth. From $-10$
to $-4 ~{\rm rad~m^{-2}}$ there is a northwest to southeast gradient of emission, which starts as 
a small-scale feature and builds up to an extended northeast-southwest structure. The mean
surface brightness of this emission is $1.8~{\rm mJy~PSF^{-1}~RMSF^{-1}}$. From $-4$ to $-0.5 ~{\rm rad~m^{-2}}$  there is 
diffuse emission with patchy morphology and a mean surface brightness of $5.4~{\rm mJy~PSF^{-1}~RMSF^{-1}}$, and it shows
a gradient in the same direction as the feature at more negative Faraday depth but is less prominent. Around $+0.5~{\rm rad~m^{-2}}$, polarized emission is detected over the full primary beam. At more positive values  it becomes patchy and 
fades away towards  $+10~{\rm rad~m^{-2}}$. We note a conspicuous, stripy morphological pattern
of diffuse emission oriented North-South in the Eastern part of the image. This structure 
is visible from $0$ to $+4~{\rm rad~m^{-2}}$. A few representative Faraday spectra of these features are shown 
in Fig.~\ref{fig:Fspec}. Following \citet{schnitzeler09} we show a map of the highest peak of Faraday depth spectra 
at each spatial pixel in Fig.~\ref{fig:RMmax}. On the same figure we also show a map of the RM value of each peak.

In contrast to polarized intensity, which reflects the amplitude of polarized emission, Stokes Q,U 
reflects the morphology of both its amplitude and its polarization angle. Therefore, 
Stokes Q,U images show even more striking morphological patterns and structures in polarization (Fig.~\ref{fig:PI}).
Particularly noteworthy are the faint linear features at Faraday depths  around $+1~{\rm rad~m^{-2}}$ that are visible 
on the Eastern side of the Stokes Q,U images. They point to large gradients in the polarization position angle in the direction orthogonal to the  
long axes. 

We also compute the total polarized intensity at each pixel by integrating the polarized intensity RM cube along Faraday depth. 
The integral is given by  \citep{brentjens11}:
\begin{equation}\label{eq:intP}
PI=\frac{1}{B}\sum_{i=0}^{n}\left(|PI(\Phi_i)|-\overline{n_{PI}}\right),
\end{equation}
where $B$ is the area under the restoring beam of RMCLEAN \citep{heald09} divided by $\Delta\Phi=|\Phi_{i+1}-\Phi_{i}|$ and $\overline{n_{PI}}$
is the mean value of the polarized intensity in the RM cube in regions where no signal is present. Assuming that the noise distributions of Q,U RM cubes have equal $\sigma$ and zero mean, and are uncorrelated, then the mean value of the PI is $\overline{n_{PI}}=\sigma\sqrt{\frac{\pi}{2}}$. 

We have not attempted to deconvolve our RM cubes for the effects of the side lobes of the RMSF  (see Fig.~\ref{fig:rmsf}). The sidelobes are small and the S/N is generally so low that this would have had very little effect on the images.  Moreover, deconvolution in Faraday space is a nontrivial operation
with many uncertainties due to missing structure in $\lambda^2$-space (see \citep{brentjens05}.

\begin{figure*}[tb]
\centering \includegraphics[width=.45\textwidth]{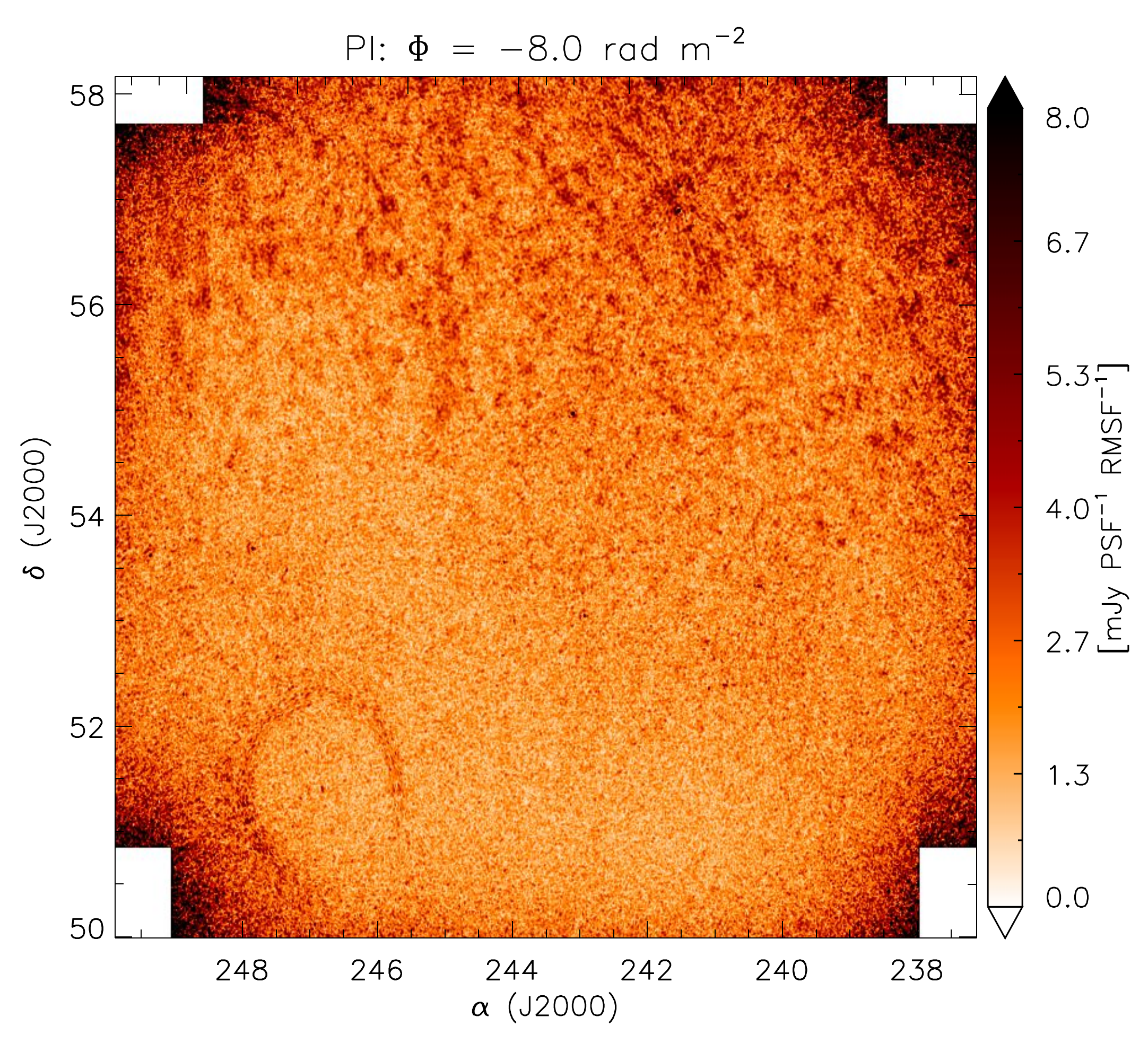}
\centering \includegraphics[width=.45\textwidth]{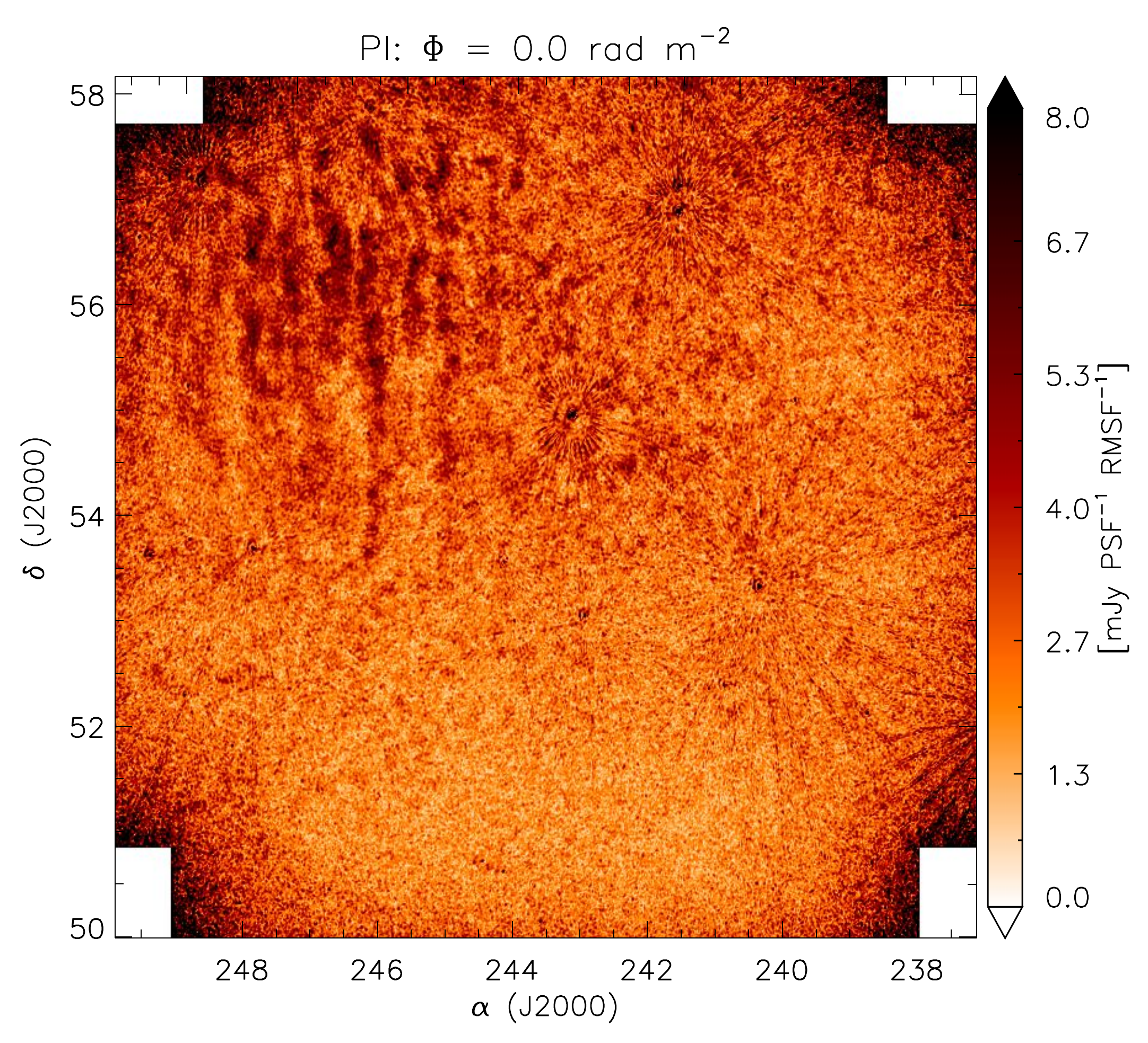}
\centering \includegraphics[width=.45\textwidth]{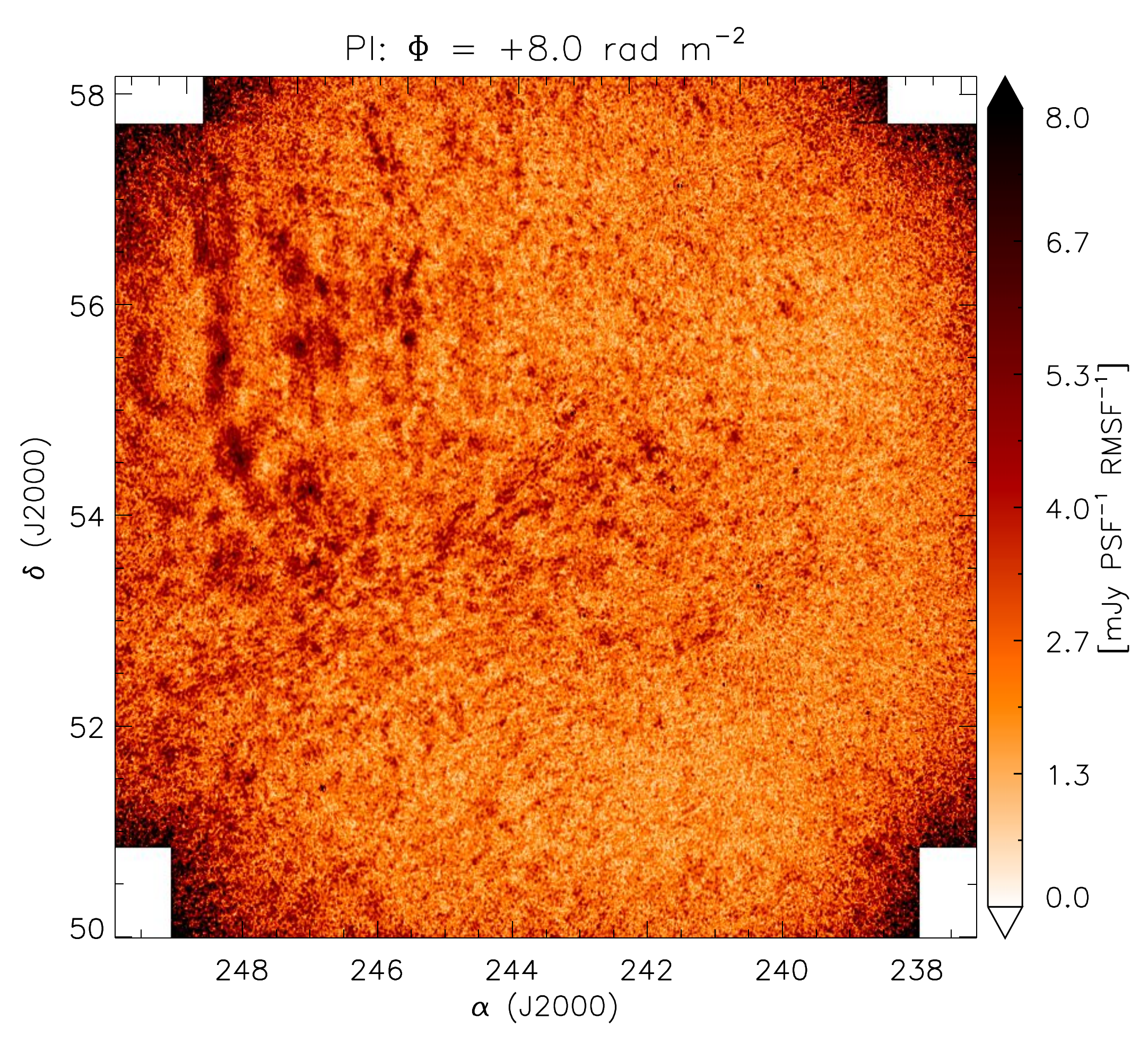}
\centering \includegraphics[width=.45\textwidth]{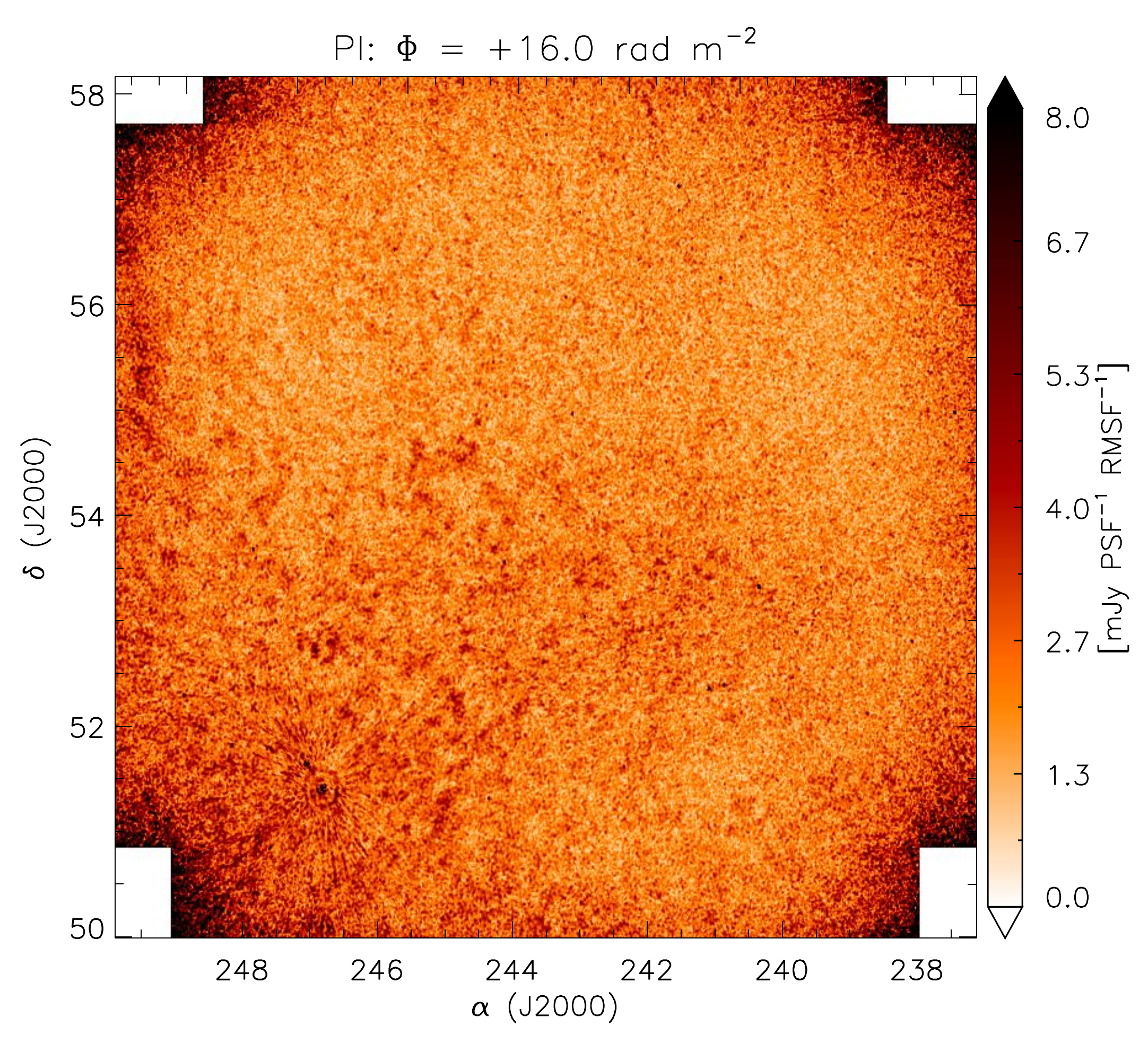}
\caption{Images of the ELAIS-N1 region in polarized intensity (PI) observed with the 28 WSRT mosaic pointings at 350 MHz (based on preliminary calibration and analysis). The images are given at Faraday depths of -8, 0, +8, +16~${\rm rad~m^{-2}}$. Images are $8.2^\circ\times8.2^\circ$ in size with a PSF of $2'\times3'$ and the noise level is $0.4~{\rm mJy~PSF^{-1}~RMSF^{-1}}$. The images have not been corrected for polarization noise bias.}
\label{fig:WSRT_P}
\end{figure*}

To calculate parameter $B$ we use the area under the RMSF instead of the restoring beam of the RM-CLEAN. Fig.~\ref{fig:intP} shows an image of the integrated polarized intensity. Most of the sources visible in the image are instrumentally polarized and appear around $\Phi=-1~{\rm rad~m^{-2}}$, which corresponds to $\Phi=0~{\rm rad~m^{-2}}$ in RM cubes not corrected for ionospheric Faraday rotation. We mask intrinsically/instrumentally polarized sources in the RM cubes to be able to study the properties of diffuse emission. The average surface brightness of diffuse emission is $3.5~{\rm mJy~PSF^{-1}}$.

\section{Discussion}\label{sec:discuss}
\subsection{Properties of Galactic polarized emission}
Detected diffuse emission in polarization that spans over $\Delta\Phi\approxeq 15~{\rm rad~m^{-2}}$ is evidently Galactic. It has a brightness temperature on average of $\sim4~{\rm K}$\footnote{At 160~{\rm MHz} and a PSF of $3.4'\times3.1'$, $1~{\rm mJy~PSF^{-1}}$ corresponds to a brightness temperature of $1.3~{\rm K}$.}. 
Its morphological structures and Faraday depth range are similar to polarized emission seen at $350~{\rm MHz}$ in a number of fields at mid/high Galactic latitudes \citep[e.g.][de Bruyn \& Pizzo, submitted to A\&A]{wieringa93, haverkorn03a, haverkorn03b, pizzoPhD}.  For example, the patchy structures and ``canals'' seen in polarized intensity (see Fig.~\ref{fig:PI}) and sharp stripy patterns in Stokes Q,U (see Fig.~\ref{fig:PI}) are quite similar to the morphological features seen at 350 MHz in the direction of A2255 \citep{pizzoPhD}, which is at comparable Galactic latitude ($b_{\rm A2255}=+35^\circ$ and $b_{\rm ELAIS-N1}=+44^\circ$). The higher spatial and RM resolution of the LOFAR observations reveal canals and filaments (i.e. at $\phi=-2.5~{\rm rad~m^{-2}}$) more pronounced than previous observations of fields at the same frequency and at high Galactic latitude \citep{bernardi10} and in a large fraction of the Southern sky \citep{bernardi13}. 

\begin{figure*}[tb]
\centering \includegraphics[width=.33\textwidth]{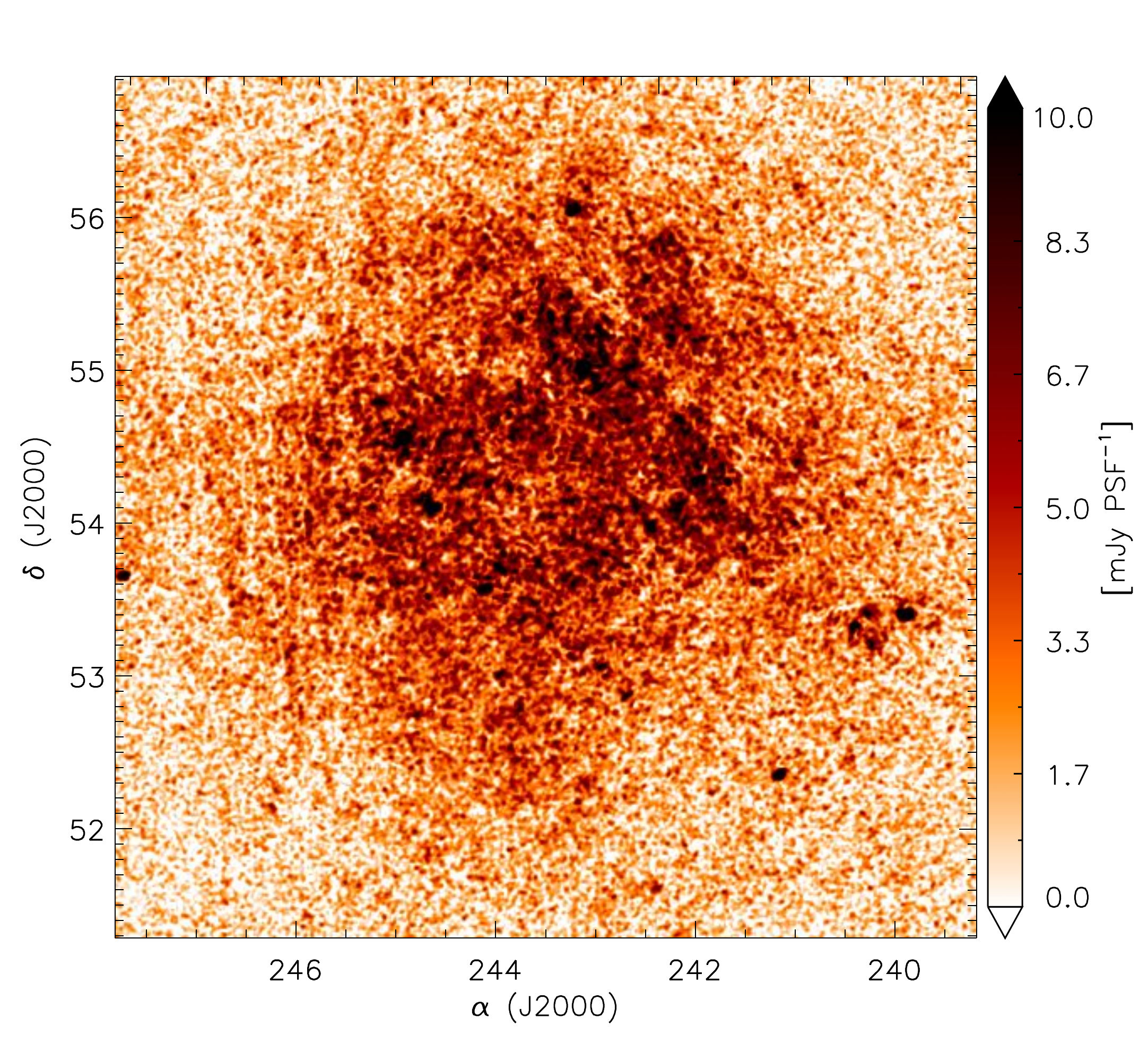}
\centering \includegraphics[width=.33\textwidth]{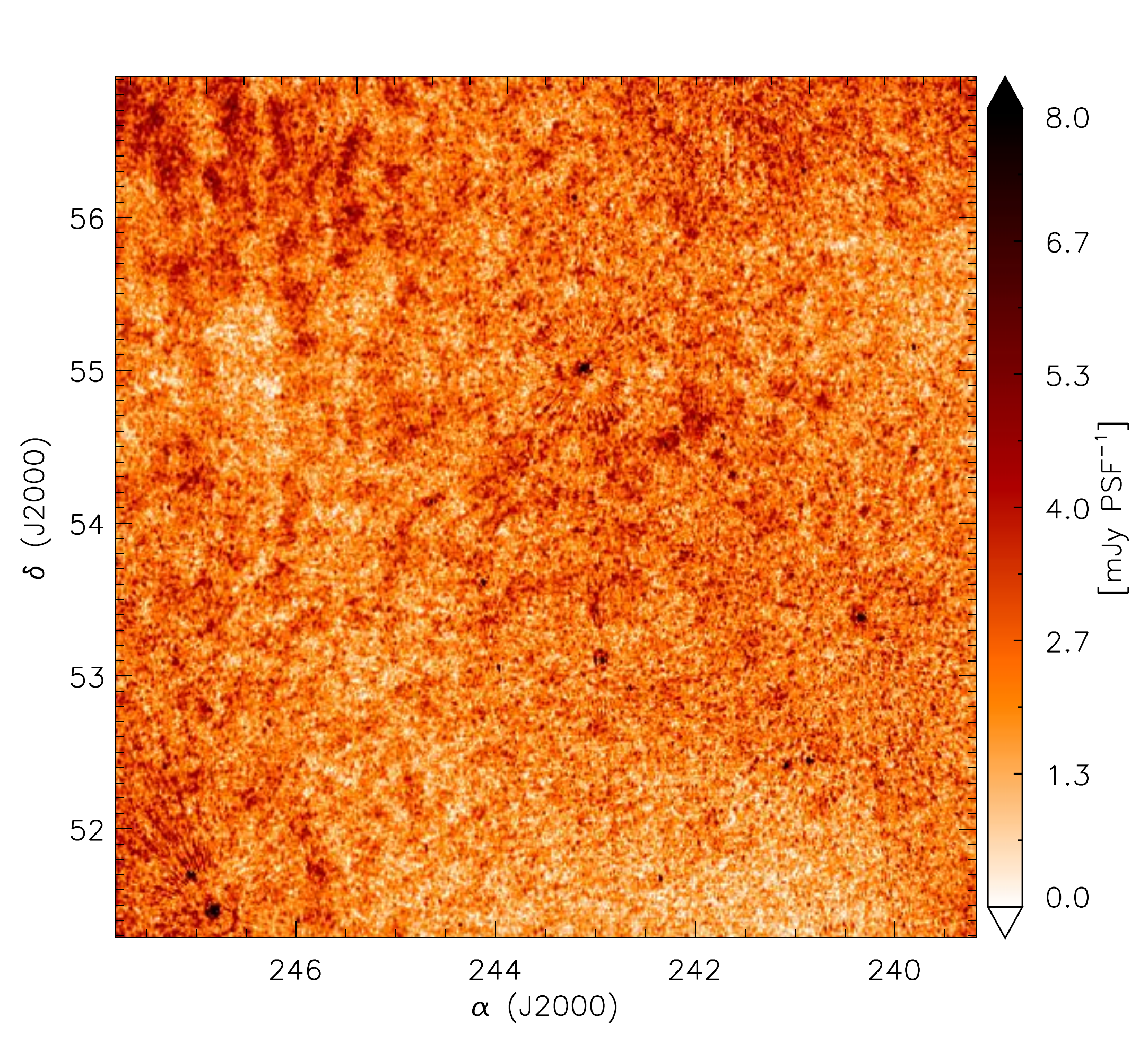}
\centering \includegraphics[width=.33\textwidth]{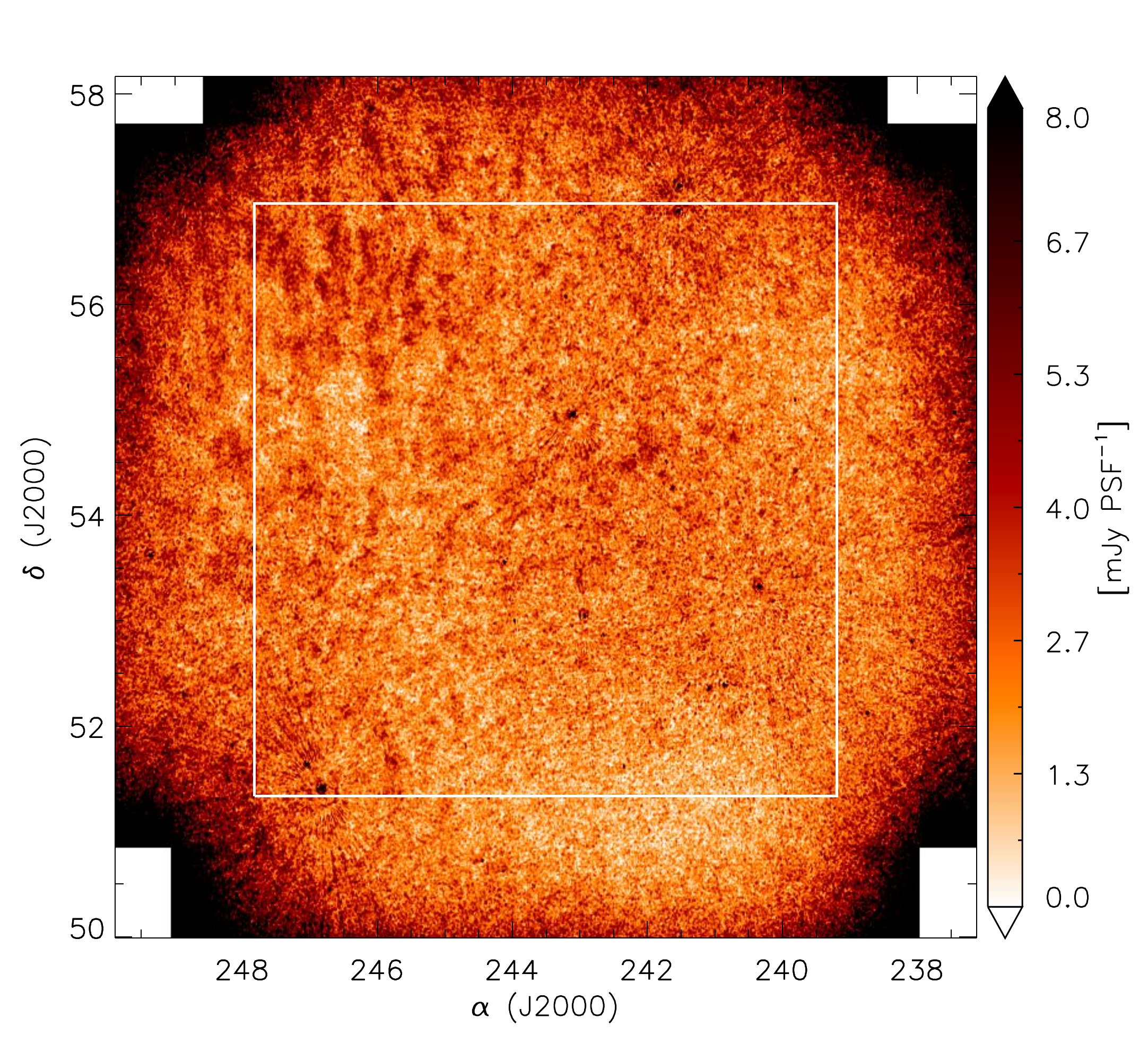}
\caption{Total polarized intensity image of the ELAIS-N1 region integrated along Faraday depth at $160~{\rm MHz}$ (observed with a single LOFAR pointing; left panel) and at $350~{\rm MHz}$ (observed with the 28 WSRT mosaic pointings; right panel). A region observed with the LOFAR is smaller than a region observed with the WSRT. Hence, the image at $160~{\rm MHz}$ measures $5.7^\circ\times5.7^\circ$ in size and the image at $350~{\rm MHz}$ measures $8.2^\circ\times8.2^\circ$ in size. The white box marks a common region of the two images. A middle panel shows a part of the WSRT image within the white box. A PSF is $3.4'\times3.1'$ at $160~{\rm MHz}$ and $2'\times3'$ at $350~{\rm MHz}$.  A flux of $1~{\rm mJy~PSF^{-1}}$ corresponds to a brightness temperature of 0.46 K at $350~{\rm MHz}$,  respectively 1.3 K at $160~{\rm MHz}$.  The strongest emission in the WSRT 350~MHz  image occurs in the North-Eastern part. Unfortunately the size of the LOFAR station beam at 160~MHz prevents us from detecting any corresponding emission at 150~MHz.}
\label{fig:intP}
\end{figure*}

We also estimate the fractional polarization by dividing the polarized intensity, integrated over all Faraday depths, by the $408~{\rm MHz}$ total intensity map \citep{haslam81, haslam82} scaled to $160~{\rm MHz}$. The spectral index between 45 and $408~{\rm MHz}$ of Galactic synchrotron emission in this region is $\beta=-2.6$ \citep{guzman11}. If we scale the brightness temperature of $\sim25~{\rm K}$ from $408~{\rm MHz}$ to $160~{\rm MHz}$, we deduce a brightness temperature of $285~{\rm K}$ in total emission. The observed polarization levels therefore imply a polarization of $\approx1.5\%$.

The maximal intrinsic fraction of polarization depends on the energy spectral index, $p$, of cosmic-ray electrons that interact with the Galactic magnetic field and produce most of the synchrotron emission in our Galaxy. The expected intrinsic fraction of Galactic polarization is at most $\Pi (p=-2.1)=(|p|+1)/(|p|+7/3)=69.9\%$ \citep[e.g.][]{sun08}. This value is much higher than what we observe.  To reach this maximum percentage however requires a uniform magnetic field in the emitting region, a situation that is rarely achieved in a physically deep emitting region. In addition, the emitting region may not be Faraday thin, leading to a sharp, frequency-dependent reduction in the emerging polarized flux, which can not be recovered using RM synthesis. 

As mentioned in Sect.~\ref{sec:intro}, the ELAIS-N1 region was also observed at 350~{\rm MHz} as a part of a WSRT continuum legacy survey. In that survey  we observed an area of $64~{\rm deg^2}$ with a 28-pointing mosaic. A preliminary calibration and analysis of part of the 350~{\rm MHz} data show large-scale emission, located mostly in the North-East part of the mosaic at Faraday depths ranging from $-10~{\rm rad~m^{-2}}$ to $+10~{\rm rad~m^{-2}}$ (see Fig.~\ref{fig:WSRT_P}). Fig.~\ref{fig:intP} shows this emission in total polarized intensity integrated along Faraday depth. In the North-East region where the most prominent  features were detected at 350~{\rm MHz}, we do not have enough sensitivity in the current LOFAR data to make a meaningful comparison. We can only see that these features have the same orientation as features detected at $\Phi=+0.5-1.5~{\rm rad m^{-2}}$ in the East part of the LOFAR images (see Fig.~\ref{fig:PI}). However, there are no clear traces of the prominent morphological features detected at the LOFAR frequencies in the central part of the WSRT mosaic (see Fig.~{fig:intP}).
 
The difference in observed emission between two frequencies can be attributed to many instrumental and astrophysical effects: (i) lower sensitivity and poorer resolution in Faraday space at $350~{\rm MHz}$; (ii) a complex distribution of emitting and Faraday rotating structures along the line of sight with variable Faraday depths; and (iii) depolarization that is more prominent at lower radio frequencies than at higher frequencies. If we were to scale the observed polarized emission at LOFAR frequencies to $350~{\rm MHz}$ we would expect to see emission levels of at least $0.9~{\rm K}$ or higher,  assuming a spectral index of $\beta=-2.6$.  The noise  in our preliminary RM cubes at $350~{\rm MHz}$ is $\sim0.46~{\rm K}$. Thus at $350~{\rm MHz}$, we are able to detect just the brightest peaks of emission observed at LOFAR frequencies.

We also note that the resolution in Faraday depth at $350~{\rm MHz}$ is an order of magnitude worse than that at LOFAR frequencies. It is possible that multiple Faraday thin structures ($\Delta\Phi<\delta\Phi_{350~\rm MHz}$) detected in the LOFAR low-frequency images will decorrelate  when we observe them with a much broader RMSF. To test this we have also generated RM cubes at LOFAR frequencies using an RMSF that has a resolution of $\delta\Phi_{350~\rm MHz}$. A new image of total polarized intensity integrated along Faraday space shows only 45\% correlation with the image in Fig.~\ref{fig:intP}. This means that $\sim55\%$ of the polarized emission detected with the full range of LOFAR frequencies will remain undetectable at $350~{\rm MHz}$.

The underlying distribution of synchrotron-emitting and Faraday rotating structures is known to be very complex. For a detailed discussion we refer to de Bruyn \& Pizzo (\textit{submitted to A\&A}),  who carried out a detailed analysis of the Galactic foreground structures in the direction of the cluster Abell 2255. To incorporate the LOFAR and WSRT RM cubes, taken in two wide but discontinuous  frequency ranges, into one physical picture  will probably require  a 3D-model and a full radiative transfer analysis.  This is beyond the scope of this preliminary analysis and we will leave it for future work once we have fully analysed the WSRT $350~{\rm MHz}$ data and incorporated deeper LOFAR observations of the ELAIS-N1 field. To perform such modelling will probably require data over the full frequency range.

\subsection{Foreground emission in the LOFAR-EoR experiment}
One of the major astrophysical challenges for the EoR experiments is the extraction of the cosmological 21-cm signal from the prominent astrophysical foregrounds \citep[e.g.][and references therein]{jelicPhD}. The extraction is usually done in total intensity along frequency. The cosmological 21 cm signal is essentially unpolarized and fluctuates along frequency. The foregrounds are smooth along frequency in total intensity and might show fluctuations in polarized intensity. Therefore, the EoR signal can be extracted from the foreground emission by fitting out the smooth component of the foregrounds along frequency, as shown for the LOFAR case by \citet{jelic08, harker09, chapman12, chapman13}. 

The LOFAR radio telescope has an instrumentally polarized response \citep{haarlem13}, which needs to be calibrated. If calibration of the instrument and modelling of and correction for the beam polarization is not accurate, the Stokes Q,U signals can leak to Stokes I and vice versa. Leaked polarized emission can introduce frequency dependent signals that can mimic as cosmological 21-cm signal, making extraction and analysis more demanding. This was addressed for the first time through simulations by \citet{jelic10}. \citet{geil11} 
showed, in the case of a simple foreground model and a single thin Faraday screen, how RM synthesis may be used to separate the cosmological signal from the leaked polarized foregrounds.  How to deal with spatially varying instrumental polarization leakage and complex Faraday spectra has not yet been addressed. 

The ELAIS-N1 region shows very complex Galactic polarized emission of $\sim4~{\rm K}$. Assuming a residual leakage of  0.1--0.2\%, constrained by current data analysis tools,  we may still expect error signals of $\sim4-8~{\rm mK}$ in Stokes I. Current noise levels in Stokes I images are still higher than this. Stokes I images are also confusion limited and dominated by point sources. We need to subtract as many sources as possible, using \texttt{SAGEcal}, to lower the noise in the images and hence be able to analyse the leakage of polarized diffuse emission and to test RM synthesis as a potential method for dealing with the leakages. We also need to model the LOFAR beam to high accuracy. All of this goes beyond the main purpose of this paper but we will address it in future work.

\section{Summary and Conclusions}
We have presented results from a LOFAR HBA observation of the ELAIS-N1 region, taken as a part of commissioning activities to characterise the foregrounds in the LOFAR-EoR observing fields and the LOFAR performance. We have detected polarized diffuse emission over a wide range of Faraday depths, ranging from $-10$ to $+13~{\rm rad~m^{-2}}$. 
The average brightness temperature of this polarized emission is $\sim4~{\rm K}$. This is much more than it was anticipated on the basis of earlier WSRT \citep[e.g.][]{bernardi09,bernardi10,pizzo11} and GMRT observations \citep[e.g.][]{pen09} in the same frequency band. First results from MWA were ambiguous in the observed intensity and morphology \citep{bernardi13}. The wide range of morphological features detected in ELAIS-N1 field at LOFAR frequencies are reminiscent of those observed in the Galactic polarized emission at 350 MHz \citep[e.g. in the direction of A2255,][]{pizzoPhD}.

The ELAIS-N1 region was also observed at $350~{\rm MHz}$  with the WSRT. A preliminary analysis of these data reveal a large-scale gradient of Galactic polarized emission, in the upper left part of the mosaic and at Faraday depths ranging from $-10~{\rm rad~m^{-2}}$ to $+10~{\rm rad~m^{-2}}$.  A detailed comparison between the signals observed with LOFAR  and the WSRT is not yet possible. The most significant correlation between the patterns observed in the two frequency bands are the vertical stripy patterns seen on the East and North-Eastern side of the images. However, the S/N in the WSRT data is too low, and the region where the most prominent signals are seen at 350~{\rm MHz} falls outside the primary beam of the current LOFAR observations,  to speculate about the nature of this correlation.
 
The presence of intrinsic polarization signals at levels of several K with complicated structure over a wide range of  Faraday depths will seriously effect  epoch of reionization experiments. The instrumental polarization of LOFAR  will have to be calibrated to a small fraction of a percent to limit leakage of polarization signals to levels of a few mK.   We will return to this problem when more sensitive  observations obtained in LOFAR Cycle 0 will be analysed.

Even though the presented results have a preliminary nature, they show the potential of low frequency polarimetry with LOFAR to study the ISM at high Galactic latitudes.  \citet{iacobelli13} showed how it is possible with LOFAR to study interstellar turbulence through fluctuations in synchrotron emission in a special low Galactic latitude region,  known as the  Fan, which has long been known for its exceptionally bright polarized emission.   Combining these two results, we can conclude that the wide frequency coverage and high angular resolution make LOFAR an exquisite instrument for studying Galactic polarized emission at a resolution of $\sim1-2~{\rm rad~m^{-2}}$ in Faraday depth. In combination with detailed simulations they will permit us to study the underlying 3D distribution of synchrotron emitting and Faraday rotating structures and constrain the properties of interstellar medium, turbulence and magnetic fields.

\begin{acknowledgements} 
We thank an anonymous referee for useful comments that improved the manuscript. VJ would like to thank the Netherlands Foundation for Scientific Research for financial support through VENI grant 639.041.336. CF acknowledges financial support by the {\it ``Agence Nationale de la Recherche''} through grant ANR-09-JCJC-0001-01. The Low-Frequency Array (LOFAR) was designed and constructed by ASTRON, the Netherlands Institute for Radio Astronomy, and has facilities in several countries, which are owned by various parties (each with their own funding sources), and that are collectively operated by the International LOFAR Telescope (ILT) foundation under a joint scientific policy. The WSRT is operated by ASTRON/NWO. 
\end{acknowledgements} 

\bibliographystyle{aa}
\bibliography{reflistElais}
\end{document}